\documentclass[useAMS,usenatbib, usegraphicx]{mn2e}

\title[Star Formation Quenching]{How Is Star Formation Quenched in Massive Galaxies?} 

\author[Gabor et al.]{
J. M. Gabor,$^{1}$\thanks{Email:jgabor@as.arizona.edu}
R. Dav\'e,$^{1}$
K. Finlator$^{1,2,3}$ \&
B. D. Oppenheimer$^{1,4,5}$ \\ 
$^{1}$University of Arizona, 933 N. Cherry Ave, Tucson, AZ 85721\\
$^{2}$University of California, Santa Barbara Physics, Santa Barbara, CA 93106\\
$^{3}$Hubble Fellow\\
$^{4}$Leiden Observatory, Leiden University, PO Box 9513, 2300 RA Leiden, the Netherlands\\
$^{5}$NOVA Fellow\\
}

\begin{document}

\date{11 Jan 2010}

\pagerange{\pageref{firstpage}--\pageref{lastpage}} \pubyear{2010} 

\maketitle
\label{firstpage}

 \begin{abstract}
The bimodality in observed present-day galaxy colours has long been
a challenge for hierarchical galaxy formation models, as it requires
some physical process to quench (and keep quenched) star formation
in massive galaxies.  Here we examine phenomenological models of
quenching by post-processing the star formation histories of galaxies
from cosmological hydrodynamic simulations that reproduce observations
of star-forming galaxies reasonably well.  We consider recipes for
quenching based on major mergers, halo mass thresholds, gas temperature
thresholds, and variants thereof.  We compare the resulting simulated
star formation histories to observed $g-r$ colour-magnitude diagrams
and red and blue luminosity functions from SDSS.  The merger and
halo mass quenching scenarios each yield a distinct red sequence and
blue cloud of galaxies that are in broad agreement with data, albeit
only under rather extreme assumptions.  In detail, however, the
simulated red sequence slope and amplitude in both scenarios are
somewhat discrepant, perhaps traceable to low metallicities in
simulated galaxies.  Merger quenching produces more massive blue
galaxies, earlier quenching, and more frosting of young stars;
comparing to relevant data tends to favor merger over halo mass
quenching.  Although physically-motivated quenching models can
produce a red sequence, interesting generic discrepancies remain
that indicate that additional physics is required to reproduce the
star formation and enrichment histories of red and dead galaxies.
\end{abstract}
\begin{keywords}
galaxies:evolution -- galaxies: luminosity function
\end{keywords}

\section{Introduction}

Galaxies in the local universe broadly fall into two main categories:
blue star-forming disks typically found in lower-density environments,
and red and ``dead'' ellipticals found in galaxy clusters.  Some
of the earliest extragalactic research revealed the two main
morphological categories \citep[e.g.][]{hubble26}, and later work
uncovered relationships between morphologies and larger scale
environments \citep{oemler74,davis76,dressler80}.  More recent work
employing data from large surveys has solidified a picture of galaxy
bimodality, showing distinct divisions in galaxy colour \citep{strateva01,
baldry04, balogh04}, which serves as a proxy for star formation
rate per unit stellar mass.  Surveys such as the Sloan Digital Sky
Survey (SDSS) have characterized the statistics of red and blue
galaxies to remarkable precision.  Red spheroids dominate the massive
galaxy population (with stellar masses above $\sim10^{10.5} M_{\sun}$),
while less massive galaxies tend to be star-forming disks
\citep{kauffmann03_sf_stellarmass}, with luminosity functions showing
that the relative fraction of red galaxies grows with luminosity.
Surveys pushing to higher redshifts have robustly demonstrated that
galaxy bimodality exists out to redshift $z=1$ \citep{bell04,
weiner05, willmer06}, and possibly to $z=2$ and beyond \citep{kriek08,
taylor09, brammer09}.  The properties and evolution of massive
galaxies is one of the best-studied observational areas of extragalactic
astronomy~\citep[e.g.][]{faber07}.


\subsection{Why Is Quenching Needed?}

Despite the accumulating wealth of data, the physical origin of the
bimodality in galaxy properties remains poorly understood.  Red
passive galaxies evolve from blue star-forming ones, in the sense
that galaxies dominated by old stars today must have built up their
stellar mass through star formation at early epochs.  Early theories
of galaxy formation proposed that primordial clouds of gas collapsed
under gravity to form rotating discs that harbor star formation
\citep[e.g.][]{eggen62}.  In hierarchical models, smaller galaxies
then merge to build up larger ones \citep[e.g.][]{white78}.  Such
mergers are thought to be capable of transforming spirals into
ellipticals through well-understood dynamical
processes~\citep{toomre72,mihos96}, but it is less clear why this
also results in a colour transformation from blue and star-forming
to red and dead.  Recent efforts to measure the global mass build-up
of the red galaxy population over cosmic time strongly suggest this
type of transformation.  The total stellar mass in red galaxies has
grown by roughly a factor of two since z$\sim$1, whereas blue
galaxies have changed little \citep{bell04, faber07}.  Furthermore,
evidence suggests the transition from blue to red is rapid, $\sim1$
Gyr \citep{bell04, blanton06}.

These data call for some mechanism(s) that ``quench,'' or quickly
shut off, star formation.  Given the correlation between morphology
and colour, such mechanisms (or related ones) must also lead to
morphological transformation of disk-dominated galaxies into
bulge-dominated ones, either through disruption or disk fading and
bulge growth.  Requiring that the star formation quenching occurs
rapidly likely excludes the possibility that galaxies simply consume
all of their available fuel supply.  Even without the rapid quenching
requirement, analytic arguments suggest that the dense central
regions of hot halos should cool rapidly onto galaxies in so-called
cooling flows, which are not observed.  Furthermore, simulations
show that (barring new physics) filaments in the intergalactic
medium can also continuously feed galaxies with fresh supplies of
cold gas \citep{keres09_coldmode, brooks09}.  This implies that
feedback associated with star formation is unlikely to quench
galaxies, since red and dead galaxies have no star formation yet
must remain quenched over the majority of cosmic time.  Hence some
additional physical process is required to quench star formation.

\subsection{Mechanisms that quench star formation}

There have been a number of mechanisms proposed to provide quenching.
Each mechanism either heats gas to the point where it cannot collapse
to form stars (``preventative" feedback) or expels gas that would
otherwise form stars (``ejective" feedback)~\citep{keres09_feedback}.
In this work, we focus on quenching in massive central galaxies,
so we do not address mechanisms that may quench star formation in
small satellite galaxies such as ram pressure or tidal stripping
and ``strangulation'' \citep{gunn72, balogh00}.

One well-studied mechanism is virial shock heating of intergalactic
gas falling into a galactic dark matter halo.  As gas falls in toward
the central galaxy, its gravitational energy converts to heat.  In
halos with masses above $\sim10^{12}$ M$_{\sun}$, radiative cooling
cannot keep up with gravitational heating, and a halo of hot
virialized gas forms near the virial radius \citep{birnboim03,
keres05, dekel06}.  This hot halo shocks infalling cold gas
immediately to the viral temperature.  Although it may succeed for
$\sim$2 Gyr, virial shock heating alone cannot quench all star
formation, since the hot halo gas should cool via radiation and
condense to form stars in the centres of massive halos
\citep{birnboim07}.  Additional heating of the hot halo, perhaps by an
active galactic nucleus \citep[AGN; e.g.][]{croton06} or gravitational
heating due to clumpy accretion \citep{birnboim07, dekel08, dekel09}
could potentially prevent such cooling.  This mechanism provides a
physical explanation for the mass-dependence of the galaxy bimodality
and the exponential cutoff in luminosity functions
\citep{kauffmann03_sf_stellarmass, sobral10}.

As alternatives to virial shock heating, various types of galaxy
interactions show promise for transforming galaxies.  High-velocity
encounters between galaxies may cause minor distortions and disk
heating.  Major galaxy mergers tend to transform disks to ellipticals
\citep{toomre72,springel05_mergers_ellipticals, cox06_kinematics},
and the resulting shocks and intense star formation winds may expel or
heat cold star-forming gas through shocks or feedback from supernovae
\citep[e.g.][]{cox06_feedback}.  Simulations show, however, that
processes like supernova feedback are not enough to sufficiently stifle
star formation \citep{springel05_mergers_ellipticals}.  As with virial
shock heating, some additional energy input is required.

A popular culprit to supply the necessary energy is the supermassive
black hole thought to reside at the centre of most massive galaxies
\citep[cf.][]{kormendy95, richstone98}.  When the black hole accretes
material, the resulting AGN injects energy into its surroundings
\citep[e.g.][]{mcnamara06}, supplying the necessary heating and/or
expulsion of gas.  \citet{croton06} propose that cooling and accretion
of hot halo gas triggers low-luminosity AGN at the centres of massive
elliptical galaxies, visible as radio lobes.  This so-called ``radio
mode" feedback adds enough energy to the surrounding hot halo to
prevent further cooling and star formation.

In a merger-driven quenching scenario, \citet{dimatteo05} and
\citet{hopkins06_redgals}, among others, suggest that powerful AGN
induced by galaxy mergers dump energy into the remnants of the
gaseous galactic disk, effectively superheating and expelling it
to prevent further star formation.  This ``quasar mode" feedback
operates only in the aftermath of major mergers of gas-rich galaxies.
An additional benefit of this high accretion rate mode, for which
the phenomenon was originally investigated, is that it explains the
existence of quasars and the relationships between black hole mass
and bulge properties.

Observational evidence provides some support for a connection between
AGN and galaxy evolution.  Correlations between black hole mass and bulge
properties like mass and velocity dispersion helped motivate the concept
of AGN fueling and self-regulation via feedback \citep[][]{magorrian98,
ferrarese00, gebhardt00, tremaine02}.  Furthermore, in both the local
and distant universe, AGN have been associated with a transition in
galaxies from blue to red \citep{sanchez04, silverman08, schawinski09}
and from disk-dominated to bulge-dominated \citep{kauffmann03_agn_hosts,
grogin05, gabor09}.

A handful of alternative mechanisms have recently emerged that do
not invoke AGN.  \citet{dekel09} suggest that smooth cosmic gas
accretion provokes instabilities in a galactic disk by building up
its mass, whereas clumpy accretion suppresses instabilities via
gravitational stirring.  Clumps transfer gravitational energy to
the disk, effectively heating it and suppressing star formation.
Other mechanisms, such as cosmic ray heating \citep[e.g.][]{mathews09}
and self-annhilation of dark matter particles \citep{totani05},
have not enjoyed as ample consideration in the recent literature.

\subsection{Hierarchical models of red and dead galaxies}

Several of these quenching mechanisms have been incorporated into
semi-analytic models (SAMs) of hierarchical galaxy formation.  These
models start with pure N-body simulations of dark matter, then use
the resulting dark matter halo population and merger history to
build baryonic galaxies within them using physically-motivated
prescriptions.  SAMs enable great speed and flexibility to explore
parameter space and compare the resulting observables to data, but
the large numbers of assumptions and free parameters they employ
make interpretation a challenge.

Empirically-constrained SAMs have enjoyed success at reproducing
observations of massive galaxies at low redshift.  The SAM of
\cite{croton06} based on the Millenium Simulation
\citep{springel05_millenium} established a red sequence through
mergers and disk instabilities, but required AGN feedback to prevent
additional star-forming gas from cooling onto the most massive
galaxies.  This model successfully reproduced luminosity functions
derived from 2dFGRS \citep{madgwick02}.  \citet{bower06} improved on
the ``Durham'' SAM \citep{cole00, benson03a, baugh05}, which forms
bulges as the result of strong disk instabilities, by incorporating
black hole feedback in quasistatically cooling halos.  Along with
matching low-redshift luminosity functions and colour distributions,
the improved Durham model matches observed luminosity functions, mass
functions, and the cosmic star formation rate density at redshifts
$z<2$ \citep{pozzetti03, drory03}.  \citet{cattaneo06} improved
on the GalICS SAM \citep{hatton03} by imposing a critical halo mass
above which accretion onto galaxies is shut down, and showed that this
shutdown yields better matches to galaxy color distributions and
luminosities.  \citet{somerville08} incorporate a ``unified'' model of
AGN feedback into their SAM, including both a low-accretion ``radio
mode'' and high-accretion ``bright mode'' following \citet{sijacki07}.
This model matches physical properties of galaxies, such as the
stellar mass function, along with black hole correlations with their
host bulges.  Taking a more analytical approach,
\citet{hopkins08_ellipticals} adopt dark matter halo mass functions
from simulations, populate halos with galaxies using halo occupation
distribution models, and estimate merger rates among galaxies.  By
assuming that major mergers convert star-forming disk galaxies into
spheroidal quiescent galaxies, these models are found to match mass
functions and the integrated stellar mass density of quenched
galaxies.

Overall, the general scenario that has enjoyed the most success is
as follows \citep[cf. ][]{hopkins08_ellipticals, kormendy09}: A major merger causes a transformation from a spiral to
bulge-dominated system, while concurrently feeding a central black hole
that shines briefly as a quasar and emits a small fraction of its accreted
mass-energy back into the galaxy.  This causes a rapid truncation of star
formation by driving a large fraction of star-forming gas from the galaxy,
forming a red and dead elliptical.  After the transformation, eventually
a cooling flow forms, allowing low angular momentum gas to feed the black
hole.  This triggers a low-luminosity AGN which deposits energy into the
surrounding hot gaseous halo.  This shuts off the cooling flow, which in
turn shuts off the AGN, and the cycle restarts.  SAMs incorporating this
scenario can explain observed correlations between galaxy morphology,
colour, stellar age, bulge mass, and black hole mass.


Recently, the growth of and feedback from black holes have been
incorporated into hydrodynamic cosmological simulations \citep{sijacki07, dimatteo08,
booth09}.  At substantial computational expense, these simulations
combine with N-body dynamics the complex three-dimensional dynamics
of gas inflow and outflow responsible for establishing galaxy
properties.  Such simulations employ parameters for subgrid physical
processes associated with star formation, galactic winds, and feedback
from black holes, but bypass many additional parameters for treating
gas dynamics as are required in SAMs.

Simulations incorporating AGN feedback have showed early promise
by reproducing properties of the black hole and AGN and galaxy
populations \citep{sijacki07, colberg08, croft09, degraf09, booth09}, along
with the global cosmic star formation history \citep{schaye09}.
However, the implementation of AGN feedback is quite heuristic, and
some assumptions such as spherical input of AGN feedback energy
seem dubious at face value.  Hence like with SAMs, it is premature
to regard these models as fully physical descriptions of black hole
and galaxy co-evolution.

In this paper, we take an approach that might be considered as a
hybrid between simulations and SAMs.  We begin with galaxy star
formation histories (SFHs) taken from cosmological hydrodynamic
simulations, and then apply quenching recipes by post-processing
the SFHs to see which recipe(s) best match observations of massive
galaxies.  This retains the rapid computational flexibility of SAMs
by allowing us to explore parameter space without re-running expensive
simulations, while still employing star formation histories of
galaxies that are described by fully hydrodynamic simulations up
until quenching.  We compare our simulated quenched galaxies to
observed colour-magnitude diagrams and luminosity functions in the
local universe, focusing on the massive (and bright) galaxy population
since these are well-observed, since the largest discrepancies in
existing models are found there, and since it is where the bimodality
is most pronounced.  This approach avoids any explicit reference
to the physical mechanisms of quenching, e.g. it does not explicitly
account for growth and feedback from black holes, and instead focuses
on asking the question:  Which aspects of galaxy and halo evolution
govern quenching in massive galaxies?

Our paper is organized as follows.  \S\ref{sec.simulations} describes
our base cosmological hydrodynamic simulation employed in this work.
\S\ref{sec.post-process} details how we post-process the SFHs under
various proposed quenching mechanisms.  We then compare the resulting
galaxy properties to data as described in \S\ref{sec.observations}.
Our results, in \S\ref{sec.results}, suggest that several quenching
mechanisms can indeed qualitatively reproduce a red sequence, but
they all fail to exactly match observations in sometimes subtle
ways.  We devote much of our discussion in \S\ref{sec.failures} to
understanding the failures of these models, with an eye toward what
new aspects may be required to improve agreement.  Finally, we
summarize in \S\ref{sec.conclusion}.

\section{Simulations} \label{sec.simulations}

\subsection{Input physics}

We run simulations using a modified version of the N-body + Smoothed
Particle Hydrodynamics (SPH) code GADGET-2 \citep{springel05}
described in \citet{oppenheimer08}.  In essence, SPH is a Lagrangian
implementation of hydrodynamics where particles represent fluid
elements, and each particle's density and temperature are determined
by kernel averages over its nearest neighbors.  GADGET-2 computes
gravitational forces between particles using a tree-particle-mesh
algorithm, and employs an entropy-conserving SPH formulation to
compute pressures and densities of the gas.  The code tracks three
distinct types of particle --- dark matter, gas (or SPH), and star
--- including positions, velocities, and masses, as well as
temperatures, densities, and metallicities as appropriate.

GADGET-2 incorporates a mechanism for gas particles to spawn star
particles stochastically.  This mechanism, based on an analytic
two-phase model of the interstellar medium with cold dense clouds
embedded in a diffuse ionized gas~\citep{springel03}, operates on
scales not resolved by the simulations.  The model assumes that
each gas particle that is sufficiently dense to be Jeans unstable
contains a hot ambient medium, cold clouds, and stars which form
within the cold clouds, and the phases exchange mass, energy, and
metals via condensation and supernovae energy input \citep{mckee77}.
The resulting star formation rate (SFR) in the cold clouds is
calculated assuming a \citet{schmidt59} law.  A gas particle with
a positive SFR will spawn a star particle with a mass, $M_*$, with
some probability determined by the SFR.  Any newly spawned star
particle inherits the position, velocity, and metallicity of its
parent gas particle, whose mass is reduced by $M_*$.  We use $M_*
= 0.5 M_{\rmn{gas}}$, the initial mass of a gas particle.  On
average, a gas particle can spawn up to two star particles, but
this varies because gas particles can acquire additional gas mass
from nearby stellar evolution (see below).  \citet{springel03} have
reduced this model to a single free parameter: the characteristic
time-scale over which cold clouds convert into stars at the threshold
density, which is set to 2.1 Gyrs in order to match the observed
Kennicutt relation \citep{kennicutt98}.

Galactic outflows driven by star formation are implemented in a
similarly stochastic fashion.  Our model, first introduced by
\citet{springel03}, ejects star-forming particles from galaxies by
kicking them with a velocity $v_w$.  The probability for ejection is
set by the mass loading factor $\eta$ (i.e. the mass loss rate
relative to the star formation rate) times the star formation
probability for that particle.  The simulation used here employs our
favored momentum-driven wind scalings~\citep{murray05} as described in
\citet{oppenheimer06} and \citet{oppenheimer08}, and motivated by
observations of local winds by \citet{martin05} and \citet{rupke05}.
In these scalings, $v_w$ is proportional, and $\eta$ is inversely
proportional, to the galaxy circular velocity.  We also decouple the
winds hydrodynamically until they reach a density 0.1 times the
critical density for star formation (up to a maximum duration of
20~kpc/$v_w$) in order to mock up chimneys through which outflowing
gas can escape but that are poorly treated by the
spherically-averaging SPH algorithm.  Both outflows and star formation
are to be regarded as Monte Carlo prescriptions, in which individual
events are not by themselves meaningful; resolution convergence
tests~\citep[e.g.][]{finlator06} have shown that once 64 star
particles are formed within a galaxy, then its star formation history
is fairly stable and well-resolved.  We use this as our galaxy
(stellar) mass resolution limit.

More recent improvements in \citet{oppenheimer08} include a
sophisticated chemical enrichment model, Type Ia supernovae, and
stellar mass loss.  We track carbon, oxygen, silicon, and iron
individually, with yields taken from various works.  Type Ia SNe
rates are taken from \citet{scannapieco05}, with a prompt and delayed
compenent, and these inject energy and metals (predominantly iron)
into surrounding gas.  We also track mass loss from AGB stars~\citep[using
models of][; BC03]{bc03}, injecting metal-enriched mass (but no energy)
to the nearest three gas particles.  Generally, these ``delayed
feedback" mechanisms become more important at later epochs, which
will be less of a concern for us in this work since the quenching
of star formation in massive galaxies typically occurs relatively
early on.

The momentum-driven outflow scalings, together with hydrodynamic
decoupling, have resulted in our simulations broadly matching a wide range
of galaxy and intergalactic medium properties.  These include the chemical
enrichment of the $z\sim 2-6$ IGM~\citep{oppenheimer06, oppenheimer08},
the galaxy mass-metallicity relation~\citep{dave07, finlator08}, and
high-$z$ galaxy luminosity functions~\citep{dave06, finlator07}.  As such,
while our model is not an ab initio description of galactic outflows,
it appears to modulate the SFHs of star-forming galaxies in broad accord
with observations.  Hence our simulations provides a plausible starting
point for studying the quenching of star formation across cosmic time.

\subsection{Simulation parameters}

The simulation we employ here begins at redshift $z=99$ with 512$^3$
dark matter particles and 512$^3$ gas particles in a cube of comoving
side length $l_{\rmn{box}} = 96h^{-1}$ Mpc with periodic boundary
conditions, and is evolved to $z=0$.  The volume of this simulation
allows us to sample large galaxies in high-density environments
(with halo mass up to $\sim10^{14.5} M_{\sun}$), while also providing
the dynamic range to resolve galaxies well fainter than the knee
of the luminosity function.  With a mass resolution of 1.2$\times
10^8$ $M_{\sun}$ per gas particle, we can resolve galaxies with
stellar masses $\geq 3.8\times 10^9 M_{\sun}$.  We employ a WMAP-5 concordance
cosmology \citep{komatsu09} with $H_0 \equiv 100h = 70$ km s$^{-1}$
Mpc$^{-1}$, matter density $\Omega_m = 0.28$, baryon density
$\Omega_b=0.046$, a cosmological constant with $\Omega_{\Lambda} =
0.72$, root mean square mass fluctuation at separations of 8~Mpc
$\sigma_8 = 0.82$, and a spectral index of $n=0.96$.

\subsection{Simulation outputs and analysis tools}

We configure GADGET-2 to output simulation ``snapshots'' at 108
redshifts, starting at $z=30$ and ending at $z=0$.  The proper time
between snapshots ranges from $\sim 50$ Myr at $z \sim 6$ to $\sim
330$ Myr at $z \sim 0$.  The snapshots contain information for each
dark matter, gas, and star particle in the simulation, including
position, velocity, mass, metallicity (for gas and star particles),
density and temperature (for gas particles), and time of formation
(for star particles).  We also store auxiliary information such as
star formation rates and carbon, oxygen, silicon, and iron abundances.

From this basic information we identify and extract properties of
dark matter halos and galaxies.  We use SKID (Spline Kernel
Interpolative DENMAX; http://www-hpcc.astro.washington.edu/tools/skid.html)
to identify gravitationally-bound groups of star-forming gas and
star particles that comprise a galaxy.  Halos are identified using
a spherical overdensity (SO) algorithm.  Beginning at the most bound
particles within SKID-identified galaxies, halos are expanded
radially until they reach the virial overdensity threshold.  Smaller
halos whose centres lie within the virial radius of larger ones are
subsumed, while halos whose virial regions overlap but whose centres
lie outside each others' virial regions divvy up the particles
according which centre is closer as scaled by their virial radii.
In the end, each halo has at least one galaxy by construction, and
each particle belongs to not more than one halo.  From these
identified galaxies and halos, we calculate quantities such as
masses, star formation rates, and metallicities by summing over all
member particles.

The star particles within a SKID galaxy are tagged with formation
time and metallicity, and hence provide a series of single stellar
populations (SSPs) from which one can construct the star formation
and chemical enrichment history of each galaxy.  This allows us to
calculate galaxy spectra using the high-resolution grid in the
stellar population synthesis models of \citet{bc03}.  We then measure
photometric magnitudes by convolving the galaxy spectra with
broad-band filter curves.  The code optionally computes extinction
due to dust in the galaxy; we explore the effects of different dust
models in \S\ref{sec.dust}.  Since we mainly focus on red sequence
galaxies where dust extinction is negligible \citep[e.g.][]{lauer05}, our dust prescription
does not strongly impact our results.


\section{Post-processed quenching} \label{sec.post-process}

For each simulated galaxy we compile a star formation history to
$z=0$.  Hence we can test models for quenching by modifying that
SFH in post-processing, based on some set of prescribed rules.  This
approach allows us to assess to first order the feasibility of a
wide range of quenching mechanisms in a short amount of time, without
the additional computational costs associated with building those
mechanisms directly into the simulations.  To describe our quenching
models in detail, we first present a general explanation of the
quenching process, and then give examples of the implementation for
each quenching mechanism.

Since we determine properties like galaxy stellar masses and broad-band
luminosities from groups of simulated star particles, we can mimic the
effects of quenching simply by removing star particles from consideration
when extracting these bulk properties.  Thus, at each snapshot,
we examine the conditions under which each new star particle formed.
If these conditions match the quenching condition, then we flag that star
particle to indicate that it never should have formed.  With this flag,
we track the quenched star particle through all subsequent time steps.
Finally, when we extract a galaxy's stellar mass and luminosity (at any
particular output time step), we ignore any star particles which have
been flagged as quenched.

This method has the obvious disadvantage that any gas which forms
into a quenched star particle will be locked up in that phantom star
particle (i.e. a star particle that never should have formed), rather
than remaining in its gaseous state to be tracked dynamically by the
simulation.  Furthermore, the additional feedback energy associated with
quenching could have an impact on the surrounding gas that subsequently
forms stars, which our post-processing technique cannot account for.
Physically, one can view this material as having been heated or expelled
in such a way so that it is unable to return to a cool star-forming state
within a Hubble time, and also does not interact significantly with other
infalling gas or galaxies.  Clearly this cannot be true in detail, but
this approach can still give useful first-order insights.  Hence while
this method has the advantage of computational flexibility and speed,
a fully self-consistent quenching mechanism must be incorporated into
the simulations dynamically in order to properly assess all effects;
this is under development.

We study quenching mechanisms that have been recently considered in the
literature.  Namely, we specify conditions under which star formation
should be quenched using three different models: (i) quenching induced
by galaxy mergers, (ii) quenching by virial shock heating in massive
dark matter halos, and (iii) quenching due to the inability of gas
shocked above a critical temperature and/or expelled in a wind to cool.
For each of these mechanisms, we define a quenching condition based on
an event or property that we can extract from the simulations.  Then we
ignore star particles whose formation satisfies the quenching condition,
as described above.  We refer to this as quenching the star particle.


\subsection{Quenching via galaxy mergers} \label{sec.merger}

As mentioned above, various works have suggested
that feedback from intense star formation and AGN activity induced by galaxy
mergers can heat and expel the cold gas from merger remnants, rapidly
halting star formation.  We apply this hypothesis to the simulations
by quenching star formation in galaxy merger remnants.  To do this, we
first identify remnants of major mergers by tracing growth in galaxy
stellar mass.  For each galaxy in a given output time step $t_c$, we
identify its most massive progenitor galaxy in the most recent time step
$t_p$ ($t_c > t_p$, and these times represent the age of the universe
at the corresponding time step).  If the galaxy and its progenitor
have stellar masses $M_*$ and $M_p$, respectively, then a major merger
occurred in the last timestep if $M_* = M_p (1 + 1/r)$, where $r$ is the
critical mass ratio that separates major and minor mergers.  We use the
typical value $r=3$ \citep{dasyra06, woods06, hopkins08_ellipticals} and a more extreme $r=4$,
and also require that the remnant be well-resolved in our simulation,
with $M_*>3.84\times 10^9 M_{\sun}$.  Note that since galaxies also grow
by star formation, this merging criterion can be considered conservative
in the sense that the actual merger ratio is probably higher (i.e. $r$
is larger) than the assumed value.

Motivated by recent work suggesting that remnants of galaxy mergers
may re-form into star-forming disks \citep{springel05_mergers_spirals,
robertson06,governato07, governato09, hopkins09_gas_in_mergers},
we incorporate an additional parameter, $f_{\rmn{gas}}$.
We assume that any merger remnant with a gas fraction
$M_{\rmn{gas}}/(M_{\rmn{gas}}+M_{\rmn{stars}}) >f_{\rmn{gas}}$ re-forms
into a disk galaxy, and therefore is not quenched.  If a galaxy meets
the quenching conditions, then we set its most recent merger time,
$t_{\rmn{merge}} = t_p$.  We track $t_{\rmn{merge}}$ for all galaxies
through the evolution of the simulation.  At any given time step,
galaxies inherit $t_{\rmn{merge}}$ from their most massive progenitors
of the previous time step.  Note that in some cases, merger remnants that
are absorbed by larger galaxies can effectively lose their remnant status;
this is fairly rare except at the earliest epochs.

After identifying merger remnant galaxies, we quench any star particle
that forms in such a galaxy at a time $t_{\rmn{form}}>t_{\rmn{merge}}$
Since the simulations track the time of formation $t_{\rmn{form}}$ for
each star particle and our tools match star particles to galaxies, we can
easily identify those star particles to be quenched.   Our model assumes
that, barring a subsequent merger with a larger galaxy, a merger remnant
will never recover from whatever quenching process occurs.  That is, we
maximally halt all future star formation, ignoring the possibility that
cold gas later accretes onto the remnant to rejuvenate star formation.
A more physically accurate model might account for the initial heating
of gas in the galaxy during the merger, then permit cooling as part of
the subsequent evolution.  Some authors \citep[e.g.][]{croton06} have
proposed that heating due to an intermittent AGN prevents later cooling
and star formation; we implicitly assume that such a process operates
to keeps merger remnants quenched.

\subsection{Halo Mass Quenching} \label{sec.virial}

Some authors \citep{birnboim07, dekel08, dekel09} have proposed that
hot gas in massive dark matter halos can halt the accretion of gas onto
galaxies, cutting off the fuel for star formation.  As gas falls toward
the centre of the dark matter potential well, gravitational energy
converts to thermal energy, heating the gas to temperatures near the
virial temperature.  In halos with masses $M_h \la 10^{11.4-12} M_{\sun}$,
rapid cooling prevent the virial shock from being supported, but in halos
above that mass a stable hot gas envelope can form \citep{birnboim03,
keres05, keres09_coldmode}.  The virial shock rapidly heats any further
infalling gas, stifling accretion of cold gas onto galaxies and thus
quenching star formation.  This mechanism effectively quenches star
formation in dark matter halos with $M_h > M_{c}$, where $M_{c}\sim
10^{12} M_{\sun}$.

The X-ray emission from intracluster gas around massive galaxies
indicate that the gas is cooling at rates of tens or hundreds of
$M_{\sun}$ per year.  However, young stars and reservoirs of cool
gas where the cooling flow might be deposited have not been observed
\citep[the cooling flow problem; see][for review]{peterson06},
suggesting that an additional long-term heating mechanism acts near
the halo centre.  Low-luminosity AGN could in principle provide
this heating \citep{croton06}, but the details of the heating process
are poorly understood.  Some works suggest that AGN jets generate
pressure waves or magnetic fields within intracluster gas to
isotropize energy input \citep[e.g.][]{ruszkowski04a,ruszkowski04b,
bruggen05}, though it is not clear that such processes can operate
in more typical-sized halos.

\citet{birnboim07} suggest a scenario in which dark matter halos that
reach $M_{c}$ abruptly form a rapidly expanding shock, heating the
supply of infalling gas and quenching star formation for $\sim 2$ Gyrs.
Eventually the shock slows, the halo stabilizes, and $\sim 10^{11}
M_{\sun}$ of gas quickly cools and initiates a starburst, followed by
a long-term quiescence.  This can quench star formation in galaxies
residing in groups, but galaxies in massive clusters require additional
heating from AGN or clumpy accretion to prevent recurrent star formation.

In this paper, we explicitly avoid reference to any particular
physics of halo quenching, but are rather interested in generally
testing whether quenching based on some halo mass threshold can match
observations of passive galaxies.  We therefore adopt a simple approach
in which no stars can form in halos above $M_{c}$.  To model this
process in the simulations, we first identify all dark matter halos
with $M_h>M_{c}$ in a given output time step.  Then we quench any star
particle which formed in one of those halos within the most recent time
step.  Again, since the simulations and analysis tools track the time
of formation and corresponding dark matter halo of each star particle,
we can straightforwardly identify star particles to be quenched.
We treat $M_{c}$ as a free parameter when comparing with observations,
testing plausible values in the range $M_{c}=10^{11.5-12.5} M_\odot$.

This procedure implicitly quenches star formation in all galaxies in the
dark matter halo, including both dominant (or central) and satellite
galaxies.  Therefore, when a small galaxy falls into a massive halo,
we immediately truncate its star formation rather than allowing a slow
decline like that expected for ``strangulation''~\citep[e.g.][]{simha09}.
Although the abruptness of this truncation does not reflect realistic
galaxy infall to clusters \citep{balogh00}, we are primarily interested
in the population of massive central galaxies that dominate the bright
end of the red sequence today.  We discuss variants of this mechanism,
including quenching only central galaxies, in \S\ref{variants}.

\subsection{Quenching of hot mode and wind mode}\label{sec.tmax}

\begin{figure}
\includegraphics[width=84mm]{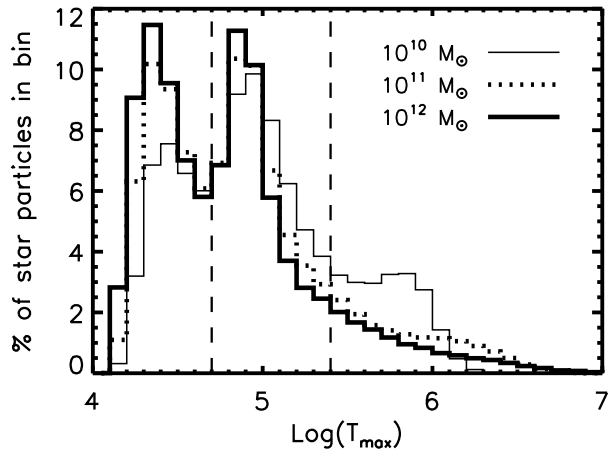}
\caption{Distribution of $T_{\rmn{max}}$ for simulated star particles
in galaxies with stellar masses in three mass bins:
$10^{10}$--$10^{10.1}, 10^{11}$ -- $10^{11.1}, 10^{12}$--$10^{12.1}
M_{\sun}$.  All masses show two distinct peaks, with a low shoulder or
tail to high temperatures.  Vertical dashed lines show our two
separate choices of the critical $T_{\rmn{max}}$ above which star
particles will be quenched.}
\label{fig.tmax_dist}
\end{figure}

Galaxies in simulations show two main paths in
density-temperature space for acquiring gas to fuel star
formation~\citep{binney77,birnboim03,keres05}:  Hot mode, in which
gas heats to near the virial temperature of the halo before cooling to
form stars; and cold mode, in which gas never gets significantly above
$\sim 10^5$~K and radiates most of its gravitational potential energy in
Lyman alpha emission~\citep{fardal01,yang06}.  Recently, \citet{oppenheimer10}
introduced recycled wind accretion, or ``wind mode", as a third accretion
path.  Wind mode constitutes gas that was previously ejected from a
galaxy, and has been re-accreted to form stars, and in our simulations
with plausible outflows models it dominates the cosmic accretion budget,
particularly in massive galaxies, at $z\la 1$.

In either the hot mode or wind mode cases, it is conceivable that
the gas should never re-accrete.  This could be owing to some
heat source that keeps hot halos energized, as hypothesized in
\citet{keres05} and \citet{dekel06}, or else it could be because there
are numerical difficulties in handling two-phase media and cold clumps
in SPH~\citep{agertz07} by which accretion is over-estimated.  Indeed,
\citet{keres09_coldmode} found that the amount of hot mode accretion
in entropy-conserving SPH is far less than in previous version of SPH,
highlighting the numerical uncertainties involved.  \citet{oppenheimer10}
also speculates that wind mode should never re-accrete, again owing
to numerical issues of cold clumps moving through hot halo gas.  It is
conceivable that, since hot mode and wind mode are preferentially more
important in larger systems, that simply removing such accretion modes
could yield a bimodality with a clear red sequence.

Our simulations track a parameter $T_{\rmn{max}}$ corresponding to the
highest temperature achieved by that particle in its history before
entering a galaxy.  The simulations stop updating $T_{\rmn{max}}$ for
gas particles ejected from galaxies via winds, even if they are heated
above the previous maximum value.  When a gas particle stochastically
spawns a star particle, the star particle inherits its parent's
$T_{\rmn{max}}$ at the time of spawning, retaining it for the
remainder of the simulation.  We thus track the maximum temperature
achieved by the gas which formed any given star particle.  To quench
hot mode, we simply assume that any gas particle heated above some
critical temperature, $T_c$, can never cool to form stars.  For this
mechanism, we simply identify and quench any star particles with
$T_{\rmn{max}} > T_c$.

Figure \ref{fig.tmax_dist} shows the distribution of $T_{\rmn{max}}$
for all star particles in galaxies in three different stellar mass
bins: $M_*=10^{10}, 10^{11}, 10^{12} M_{\sun}$.  Each distribution
shows two distinct peaks (at $\log T \simeq$ 4.4 and 4.9) and a tail
to higher temperatures.  The low-temperature peak is from gas that has
never been substantially heated, and the high-temperature peak comes
from warmer gas that then cools to form stars.  The peaks are
associated with hydrogen cooling and helium cooling, respectively
\citep[cf.][]{sutherland93}.  Mild shocks during accretion likely heat
the warm component, while the colder component somehow avoids such
shocks.  Somewhat counter-intuitively,
\citet{keres09_coldmode} found that both low and high mass galaxies
form predominantly through cold mode, and that intermediate mass
systems ($M_{\rmn{baryon}} \approx 5\times 10^{10} M_{\sun}$) have the
highest hot mode fraction.  In the high mass case, it is because large
galaxies assemble from lower-mass galaxies that formed early on mainly
via cold mode accretion.  Based on this plot, we explore critical
values of $\log T_c = 4.7, 5.4$ (shown as vertical dashed lines),
though it turns out that the results are not very sensitive to this
choice.

Since most of the high-temperature gas is heated via virial shocks,
one might expect these results to mimic those of the virial shock
heating prescription in \S \ref{sec.virial}.  In many halos, however,
cold flows of gas from the IGM (particularly at early epochs) can
penetrate the hot envelope to feed the central galaxy directly,
so the quenching is not complete \citep{keres05, keres09_coldmode}.
Our results in \S\ref{sec.results.tmax} indeed suggest that cold flows
are a significant factor in ongoing star formation in massive galaxies.

Quenching wind mode is also straightforward, since we track which
gas particles have been ejected in a wind, and this information is
passed on to spawned star particles~\citep[see][for implementation
details]{oppenheimer10}.  Certainly, it is the case that at least some
material that participated in an outflow (particularly at early epochs
when outflows were most prominent) is likely to have fallen back into
a galaxy at late times.  Here, we make the most extreme assumption that
none of it ever falls back.

\section{Observational Constraints} \label{sec.observations}

The ultimate goal of this study (and others like it) is to build a
model for galaxy evolution that matches observed global distributions
of galaxy properties.  Accordingly, we compare the results of our
simulations to well-studied quantities of the low-redshift ($z<0.1$)
galaxy population, primarily based on measured luminosities.  We focus
on colour-magnitude diagrams (CMDs) and luminosity functions (LFs).
Of course, a litany of other observations could provide additional
detailed constraints, such as luminosity- and colour-dependent
correlation functions \citep{zehavi05, weinmann06, cooper06, phleps06,
cooper07, coil08, brown08, williams09, cooper09}, properties of
post-starburst galaxies \citep{zabludoff96, quintero04, blake04,
balogh05, yang08, wild09}, the total star formation history of the
universe \citep[e.g.][]{a_hopkins06, thompson01, thompson06}, the star
formation intensity distribution of galaxies \citep{thompson02}, and
other observations of galaxy properties at higher redshifts.  Given
the challenges that even low-redshift CMDs and LFs present to our
models, we defer more detailed comparisons to future work.

We use the Value-Added Galaxy Catalog \citep[VAGC;][]{blanton05_vagc}
of the Sloan Digital Sky Survey \citep{adelman-mccarthy08,
padmanabhan08} for comparison to our simulations.  The DR6 version
of the VAGC includes k-corrected absolute magnitudes in the SDSS
ugriz + JHK bands for more than 2.6 million galaxies, including a
special low-redshift sample of $\sim$170,000 galaxies with distances
of 10 $h^{-1}$ Mpc to 150 $h^{-1}$ Mpc (redshifts roughly 0.003 to
0.05).  Because the low-redshift sample's volume is roughly comparable
to (though larger than) the volume of our simulations, we use it
(rather than the full VAGC) to provide observational constraints
on our models.  We convert the reported absolute magnitudes (which
use $h = 1$) to our preferred cosmology (with $h=0.70$) with $M_{h=.7}
= M_{h=1} + 5 \log(0.7)$.  Colours are then straightforward, and
we compute luminosity functions using the $1/V_{\rmn{max}}$ method
\citep{schmidt68} and the $V_{\rmn{max}}$ values presented in the
VAGC.

Because it effectively traces stellar mass, we plot $r$-band absolute
magnitudes in most of our plots, and we use $g-r$ colours for CMDs.
We compare observations with results from our simulation snapshots at
$z=0.025$, chosen to fall in the middle of the observed low-redshift
VAGC range, though the simulation results are insignificantly
different at $z=0$.  We refer to redshift 0.025 as $z\approx 0$, or
``low-redshift,'' throughout the remainder of the paper.

Stellar masses for galaxies in SDSS (including those in VAGC) were
determined by \citet{kauffmann03_stellarmass_measurements}, using
template fits to spectra including the effects of dust extinction.
We cross-correlate these publicly available data with the VAGC catalog
to obtain stellar masses for our comparison galaxy sample.

\section{Results}
\label{sec.results}

\begin{figure}
\includegraphics[width=84mm]{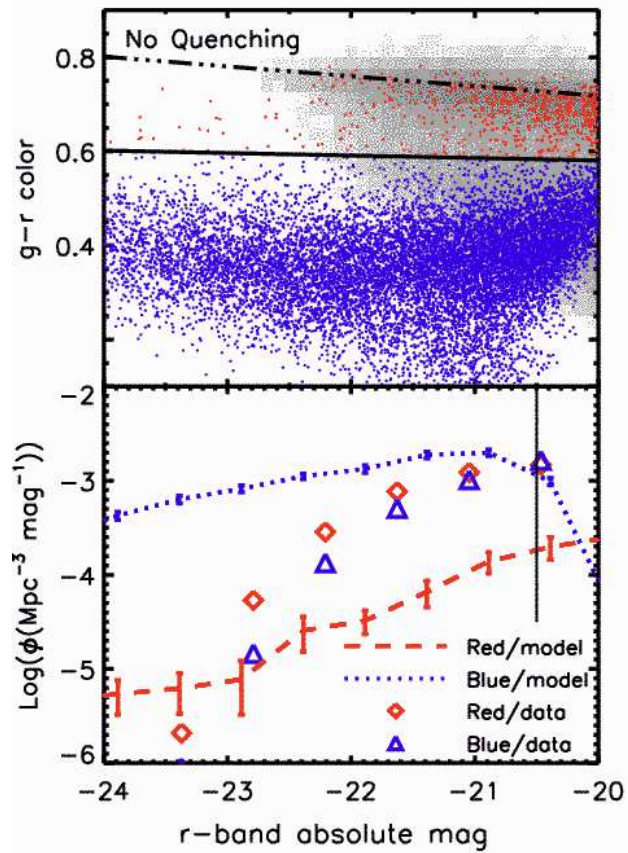}
\caption{Colour-magnitude diagram in $g-r$ vs. $r$ (top panel) and
$r$-band luminosity functions (bottom panel) for our original
simulation without quenching, compared to data from the SDSS VAGC.
The solid line separates blue and red galaxies in the simulations, and
the dot-dashed line shows a best-fit to the SDSS red sequence (see
\S\ref{sec.results.merger}).  Greyscale in top panel shows SDSS VAGC
galaxy density scaled logarithmically, and colour-coded points show
simulated galaxies.  In bottom panel, symbols represent SDSS VAGC
data, split into blue (blue triangles) and red (red diamonds)
galaxies.  Lines show simulated luminosity functions for blue (blue
dotted) and red (red dashed) galaxies.  Our mass resolution cutoff
corresponds to a diagonal envelope in the CMD at about $r=-20$, and
the cutoff affects our luminosity functions fainter than $r\sim-20.5$ (vertical solid line).
Without quenching, our simulation produces too many bright blue
galaxies, and almost no bright red ones.  }
\label{fig.noquench_cmdlf}
\end{figure}

To begin, we illustrate the current state of galaxy formation in
cosmological hydrodynamics simulations without quenching in Figure
\ref{fig.noquench_cmdlf}.  This shows CMDs (top) and LFs (bottom) for
our simulations compared with observations from SDSS.  We separate
blue galaxies from red using a solid line drawn in the CMD, and we
show a best-fit to the SDSS red sequence as a dot-dashed line (see
below).  Our mass resolution produces a diagonal envelope in the
bottom right of the CMD panel, corresponding roughly to $r\sim -20$.
This cutoff affects the blue galaxy LF fainter than $r\sim -20.5$,
although the red galaxy LF is not impacted within our plotting range,
brighter than $r=-20$.  We follow \citet{finlator06} to
estimate LF uncertainties for our simulation using the jackknife
method \citep{lupton93, zehavi02, weinberg04}.  We sample the volume
of our simulation eight times, each time excluding one octant of the
simulation and calculating the LF for the remaining $7/8$ of the total
volume.  The variance in the LF of the eight sub-samples provides an
estimate of the uncertainties due to poisson noise and cosmic
variance.  Luminosity functions of the SDSS galaxies have typical uncertainties at the 0.1 dex level \citep{blanton05_lfs}, smaller than the symbols we use in the plot.

This simulation fails to produce massive red
and dead galaxies as observed.  Almost all galaxies occupying the red
sequence regime are low-mass satellites that have been quenched
primarily via strangulation \citep[see e.g.][]{simha09}, with just a
few massive systems that are a red extension of the blue cloud,
without evidence for a distinct bimodality.

We reiterate that this simulation includes strong galactic outflows
from star-forming galaxies.  Clearly, even such fairly energetic
feedback is insufficient to quench star formation in massive galaxies.
Galaxies almost always have supplies of gas to fuel new star formation,
Galaxies accrete cold gas from the IGM through filaments of the cosmic
web \citep[cf.][]{keres09_coldmode}, and hot gas in galactic halos
may cool onto the central galaxies.  In our simulations, much of the
accreting gas today is recycled from earlier galactic winds expelled
from the galaxy \citep{oppenheimer10}.  This illustrates that some other
physical process must quench star formation in massive galaxies.


When calculating galaxy magnitudes, we neglect the effects of dust to
highlight the intrinsic red sequence, i.e. the red sequence made up
solely of galaxies with old stellar populations.  Red galaxies without
ongoing star formation contain little enough dust that we can neglect
dust effects on colours for the intrinsic red sequence.  Although dust
tends to redden star-forming galaxies from the blue cloud, observations
suggest only $\sim10$--20\% of the red sequence comes from such galaxies
\citep{bell04_dusty, brammer09}.  In \S\ref{sec.dust} we will explore
dust effects on the population of star-forming blue cloud galaxies.

In the following three sections, we present CMDs and LFs for merger-based
quenching, halo mass quenching based on a critical halo mass,
and accretion mode quenching.  We focus on the intrinsic red sequence
successfully generated in the first two models, exploring some properties
of the red galaxy populations formed.  Because of its sensitivity to dust,
we defer discussion of the blue cloud to \S\ref{sec.dust}.

\subsection{Merger quenching} \label{sec.results.merger}

\begin{figure}
\includegraphics[width = 80mm]{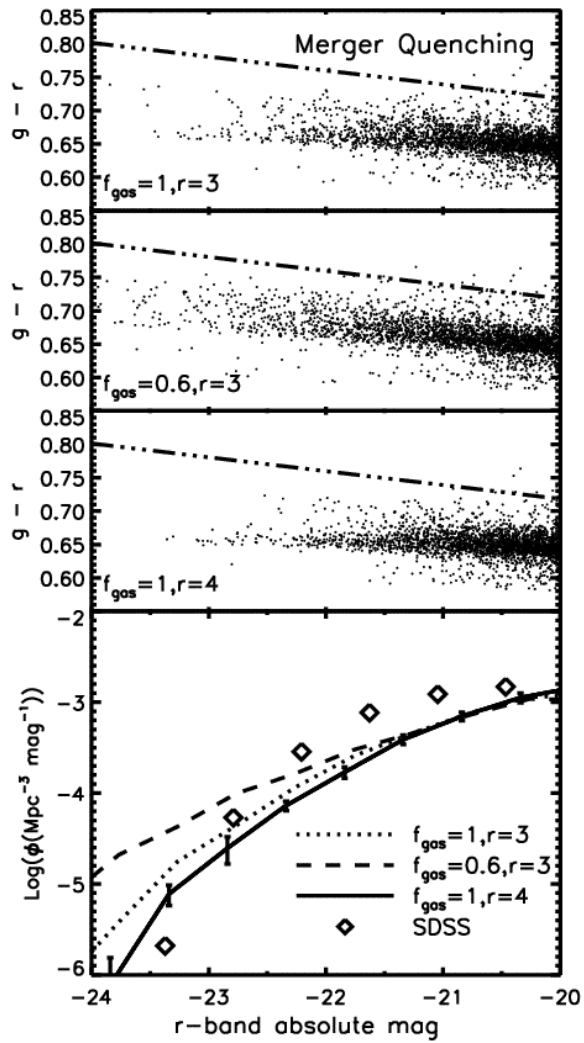}
\caption{Merger quenching colour-magnitude diagrams (top 3 panels) and
luminosity functions (bottom) for intrinsic (dust-free) red sequence
galaxies.  As indicated in the lower left of the panels, the CMDs
correspond to merger quenching with a gas fraction thresolds of
$f_{\rmn{gas}}=0.6$ and 1, above which merger remnants re-form into
star-forming disks.  In the third panel we count 1:4 $r=4$) mergers as
major, quenching star formation in the remnant.  The dot-dashed line
shows the best-fit to the SDSS VAGC red sequence.  In the bottom
panel, we show luminosity functions for red galaxies for each of the
three quenching models (lines), along with the true red sequence from
the SDSS VAGC (diamonds).  For clarity, we show representative error bars
for one simulation only.  All models overproduce bright galaxies, and underproduce
fainter ones.}
\label{fig.merger_cmdlf}
\end{figure}

\begin{figure}
\includegraphics[width = 84mm]{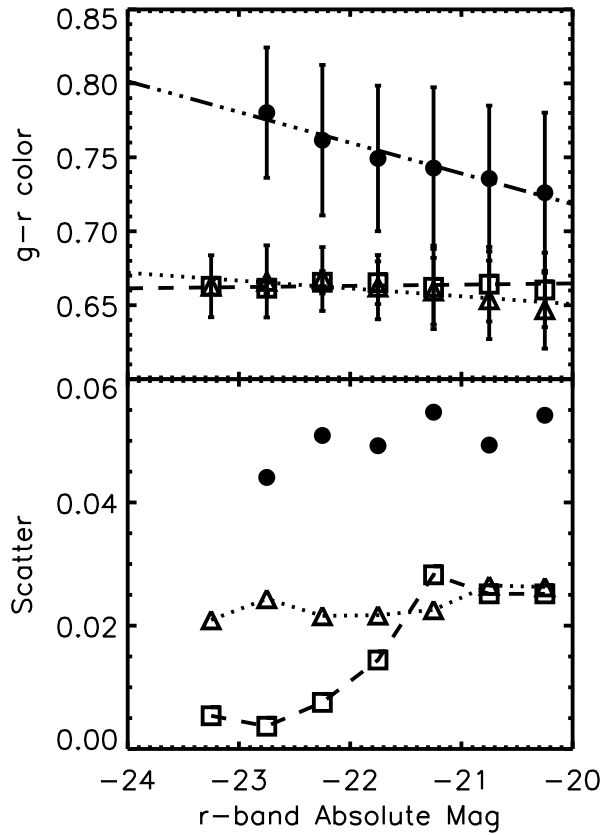}
\caption{Slope and scatter of the red sequence, for SDSS data
(circles) and two simulations: one with merger quenching
with $f_{\rmn{gas}}=1$ and $r=3$ (triangles), and another with halo mass quenching with $M_c =
10^{12}M_{\sun}$ (squares).  Our models produce red sequences that have shallow
slopes, $g-r$ colours that are too blue, and smaller scatter compared
to the true red sequence.}
\label{fig.red_seq_scatter}
\end{figure}

\begin{figure}
\includegraphics[width = 75mm]{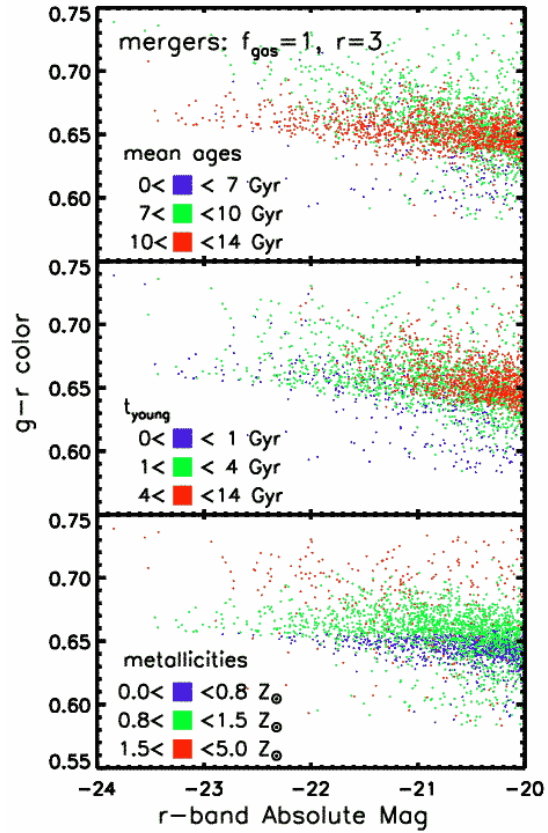}
\caption{A closer look at the CMD red sequence resulting from 3:1
mergers with $f_{\rmn{gas}}= 1$.  Galaxies are colour-coded by
$r$-band luminosity-weighted mean stellar ages (top panel);
$t_{\rmn{young}}$, the age of the youngest (most recently formed) star
particle (middle panel); and $r$-band luminosity-weighted mean stellar
metallicities (bottom panel).  For reference, solar metallicity is
$Z_{\sun}\simeq 0.012$.  Although the brightest red sequence galaxies
generally have old stellar populations, they have traces of young
stars ($t_{\rmn{young}} < 1$ Gyr).  A distinct gradient in metallicity
creates most of the spread in $g-r$ colour.}
\label{fig.red_seq_age_merger}
\end{figure}

Figure \ref{fig.merger_cmdlf} shows the results of our merger
quenching prescription with a mass ratio of 3:1 ($r=3$) with
$f_{\rmn{gas}}=0.6,1$ (top two panels) and $r=4$ with $f_{\rmn{gas}}=1$
(third panel).  CMDs in all cases exhibit a distinct red sequence
with a shallow but nonzero slope, and a tail of bright red galaxies,
qualitatively in agreement with observations.  However, the LF for
the $f_{\rmn{gas}}=0.6$ case shows a significant excess of very
bright red galaxies, which is only partly mitigated in the
$f_{\rmn{gas}}=1$ case.  Note that these are quite high gas fractions:
Disk galaxies like the Milky Way today have gas fractions well below
0.6, and even at high-$z$ one has to go to $M_*<\sim 10^{10} M_\odot$
to get typical gas fractions as high as 50\%~\citep{erb06}.  Our
results prefer {\it never} re-forming a star-forming disk once a
major merger happens.  Since hydrodynamic simulations without AGN
clearly show that a disk re-forms even with more modest gas
fractions~\citep[e.g.][]{robertson06,governato09}, this implies
that for this quenching mechanism to work, something must heat or
eject all the gas such that it not only stops forming stars at that
time, but prevents the re-formation of a gaseous disk.

In the third panel we make the model even more extreme by quenching
star formation after all 1:4 mergers, regardless of gas content.
This has only a minor impact on the resulting CMD (relative the 1:3
case).  It decreases the number of very bright galaxies, which
mitigates the discrepancy there, at the expense of worsening the
agreement for moderately bright galaxies.

The LF shows a related problem that all merger models fail to produce
a sufficiently sharp knee at a characteristic magnitude $M^*$ as
seen in the data.  The model can reasonably match the number of
galaxies at the low-luminosity end, but it over-produces bright
galaxies and underproduces galaxies at $\sim L^*$.  This is a fairly
generic problem in this scenario, present even in our most extreme
case, arising because mergers occur at a wide range of masses and
only weakly pick out a characteristic scale around $\sim L^*$.


Merger quenching (along with all our quenching models) exhibits what
we call the ``blueness problem": the red sequence is too blue by
$\sim$0.1 magnitudes in $g-r$ colour.  This corresponds to $\sim
10\%$ error in the flux ratio between the two bands.  We explore
several possible explanations for this problem in \S\ref{sec.blueness},
which likely has an origin in the overall calibration of galaxy
metallicities.  Since it occurs in all quenching models, we cannot
use this blueness problem to constrain quenching mechanisms.

To more quantitatively compare the slope, scatter, and average colour,
we perform linear fits to the simulated and real red sequences.  We
first separate the red sequence from the blue cloud using a constant
$g-r$ colour, 0.65 for real galaxies and 0.55 for simulated ones.
Various reasonable choices of this separator, including those with
nonzero slope chosen simply by eye, yield similar results.  After
dividing the absolute magnitudes into bins of width 0.5, we calculate
the mean, median, and standard deviation of $g-r$ colour within each
bin for galaxies above our separator.  We fit the median points to a
straight line, as shown in the top panel of Figure
\ref{fig.red_seq_scatter} for SDSS galaxies and our 3:1,
$f_{\rmn{gas}}=1$ merger model.  Along with highlighting the blueness
problem, the figure shows that the simulated red sequence slope is
somewhat too shallow, at least in this quenching model.  Note (in
Figure \ref{fig.merger_cmdlf}) that the model with lower
$f_{\rmn{gas}}$ produces a steeper slope, and a smaller blueness
problem, albeit failing more spectacularly in the LF comparison.

In the lower panel of Figure \ref{fig.red_seq_scatter}, we plot the
scatter as a function of absolute magnitude.  The observed scatter is
$\sim 0.05$~magnitudes, but part of that is observational
uncertainties.  \citet{cool06} estimated that the intrinsic scatter in
$g-r$ is 0.035~mag, at least for the most luminous ($>2.2L^*$)
galaxies.  This is still larger than the scatter in our simulations,
which is good because our quenching prescription makes the maximal
assumption of zero star formation since the time of quenching.
Early-type galaxies today do appear to have a small ``frosting" of
star formation~\citep{trager00,yi05}, which would tend to increase the
scatter.  However, the simulated scatter is not significantly smaller
than observed, so there is scant little room for additional scatter
from frosting.

Figure \ref{fig.red_seq_age_merger} explores the red sequence produced
in the 3:1, $f_{\rmn{gas}}=1$ merger quenching model in
greater detail.  We select this as our preferred merger quenching model
since it yields the least deviations from the observed red galaxy
LF.  The top panel shows the $r$-band luminosity-weighted mean
stellar age; the middle panel shows the simulated galaxies in colour-coded
bins of $t_\rmn{young}$, the age of the youngest star particle in the
galaxy; and the bottom panel shows the $r$-band luminosity-weighted
mean stellar metallicity.  The metallicities shown are absolute mass
fractions of elements heavier than helium, where solar metallicity is
$Z_{\sun}\approx$0.012 \citep{asplund05}.

At the brightest end, the galaxies have old stellar populations
on average.  Less massive galaxies show a wider spread in age,
and perhaps counter-intuitively, the intermediate-age galaxies are
redder than the oldest galaxies.  There are some small (i.e. $\sim L^*$)
galaxies that are fairly young, having just been quenched onto the
red sequence.  The middle panel shows that even the brightest
galaxies generally have some quite young stars ($<1$~Gyr old).  This is
because they live in the largest halos that are still assembling at
the present day, and hence they recently subsumed galaxies that were
recently forming stars (i.e. unquenched galaxies); we discuss this
further in \S\ref{sec.discussion}.  In metallicity, the red sequence
shows a gradient where the reddest galaxies are the most metal rich, and
this occurs at all luminosities.  This reflects the well-known fact from
population synthesis models that metallicity is the primary determinant
of colour in old stellar populations.  Here we can see why intermediate
age systems are redder than the oldest systems: They generally have
higher metallicities, having formed stars up until a later cosmic epoch.

\subsection{Halo mass quenching} \label{sec.results.mass}

\begin{figure}
\includegraphics[width = 80mm]{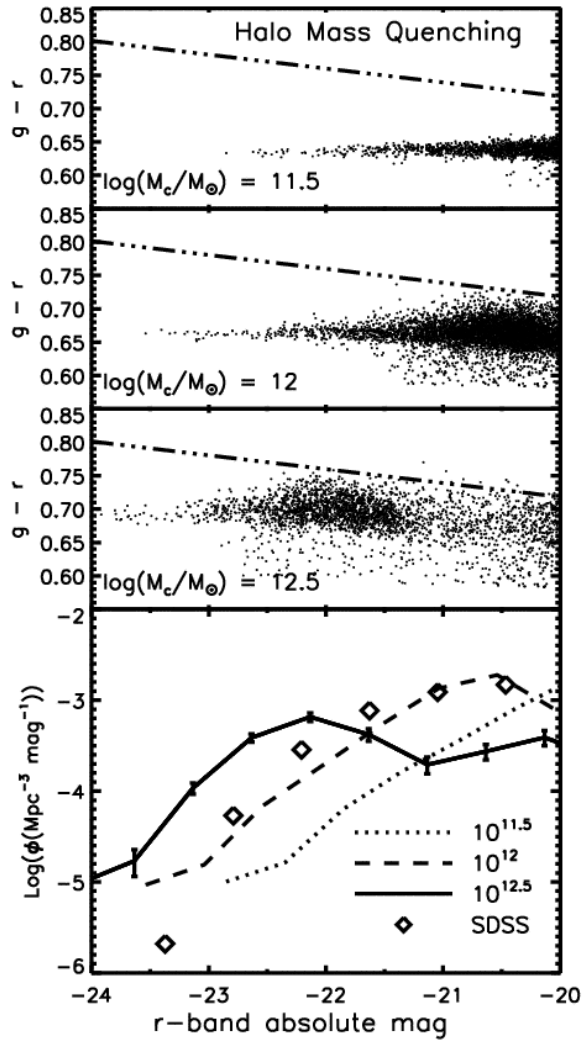}
\caption{Halo mass quenching CMDs and LFs, analogous to Figure
\ref{fig.merger_cmdlf}.  We quench star formation in halos above the
critical mass, $M_c$, for three values $M_c = 10^{11.5}, 10^{12},$ and
$10^{12.5} M_{\sun}$.  Galaxies tend to enter the red sequence in a
clump defined by a characteristic $r$-band absolute magnitude (or
stellar mass) associated with $M_c$ (e.g. $r \sim -22$ for $M_c = 10^{12.5}$).  The luminosity functions (bottom panel) show an
unobserved uptick at the brightest bins for all three cases.  The
model with $M_c = 10^{12} M_{\sun}$ yields the best match to the SDSS
luminosity function.}
\label{fig.mass_cmdlf}
\end{figure}

\begin{figure}
\includegraphics[width = 75mm]{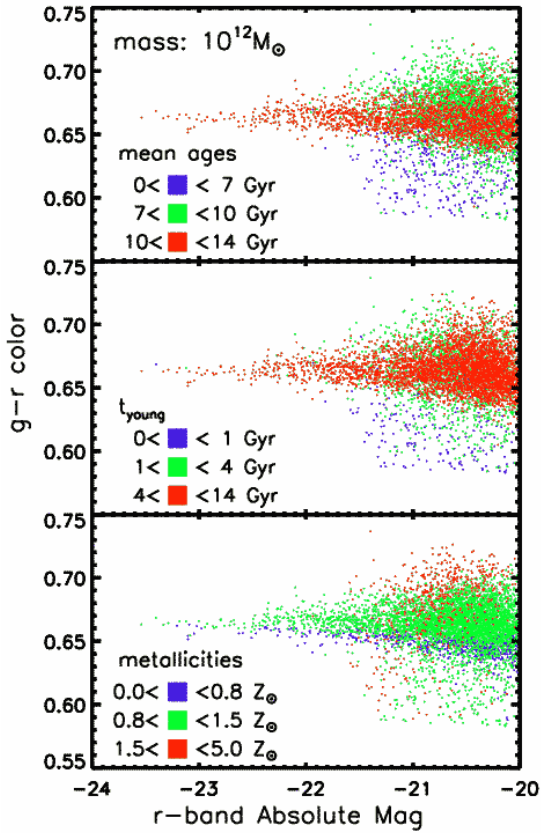}
\caption{A closer look at the CMD red sequence resulting from halo mass quenching with a critical mass of $10^{12} M_{\sun}$.  Colour
coding is the same as in figure \ref{fig.red_seq_age_merger}.
Unlike in our merger quenching model, the brightest galaxies include
no young stars.}
\label{fig.red_seq_age_mass}
\end{figure}

Figure \ref{fig.mass_cmdlf} shows CMDs and LFs for halo mass
quenching, analagous to Figure \ref{fig.merger_cmdlf} for mergers.
We show results for three values of the critical cutoff mass, $M_{c}$
= $10^{11.5}, 10^{12},$ and $10^{12.5} M_{\sun}$.  As with mergers
we obtain a distinct red sequence in qualitative agreement with
data.  The slope is slightly shallower than in the merger case,
essentially zero, as shown in Figure \ref{fig.red_seq_scatter}, and the
scatter around the red sequence drops to nearly zero for the brightest
galaxies, in contrast to merger quenching where the scatter is
independent of luminosity.  The blueness problem is present at
roughly the same level as in merger quenching.

In halo mass quenching, most red galaxies tend to clump in a relatively
small range of absolute magnitude.  For $\log M_{\rmn{c}}$=12.5 this
clumping occurs around $r=-22$ (and stellar mass $\sim$10$^{11}
M_{\sun}$), and scales with halo mass for the other $M_c$ cases.
Variations in star formation and merging histories smear out this clumping
somewhat, but this general feature remains; no such feature is seen in the
data.  The bright end of the red sequence grows large through
dry mergers after the quenching process, and faint galaxies are mostly
satellites.  The dark matter halos of central galaxies tend to achieve
the critical mass when the galaxies have a particular stellar mass, and
then move from the blue cloud to the red sequence.  Thus, at all epochs,
galaxies tend to move onto the red sequence at an effective critical
stellar mass or absolute magnitude that corresponds to the critical halo
mass.  Variations in the time of quenching and metallicities then cause
significant vertical scatter in $g-r$ colour for these clump galaxies.
At a fixed stellar mass, galaxies early in the universe (high redshift)
have lower metallicities than those later on (low redshift) \citep{erb06,
maiolino08}, so galaxies that move onto the red sequence later have
higher metallicities.  


The LF is quite sensitive to $M_c$: $M_{c}=10^{12} M_{\sun}$ fits the
luminosity functions best, with the other cases far underpredicting or
overpredicting the number of bright galaxies.  The CMD clump is evident
in the LF as well.  Even in the generally best-fitting $M_{c}=10^{12}
M_{\sun}$ case there is a tail of bright galaxies that leads to an excess
in the luminosity function at $r< -23$, just as in merger quenching.
Overall, the luminosity function produces a slightly steeper drop-off
than the merger quenching models, although a characteristic knee is
still less pronounced than in the data.

Figure \ref{fig.red_seq_age_mass} shows the red sequence in more
detail for the best-fitting $\log M_c = 12$ case, with galaxies
colour-coded by mean stellar age, $t_{\rmn{young}}$, and mean stellar
metallicity (from top to bottom).  Like in merger quenching (cf. Figure
\ref{fig.red_seq_age_merger}), we find that the most massive galaxies
have old stellar populations.  Unlike in merger quenching, these
old massive galaxies do not contain any young stars.  Each of these
galaxies, lying at the centre of a cluster, tends to merge only with
smaller galaxies that have already been quenched a while ago because
they live in the same dark matter halo above the critical halo mass.
Thus, when the central galaxies grow through accretion of satellites,
they do not obtain any stars younger than 1 Gyr.

The metallicity panel shows that the highest metallicity systems are
generally the smallest (and reddest) systems.  The large, old systems
typically have fairly low metallicities.  Qualitatively, this explains
why the red sequence slope is basically zero; the age and metallicity
gradients cancel each other.  The tight correlation between luminosity
and metallicity directly translates into a scatter in the red sequence
that goes from nearly zero at the brightest end to fairly large at the
faint end, as seen in Figure~\ref{fig.red_seq_scatter}.

\subsection{Accretion mode quenching}
\label{sec.results.tmax}

\begin{figure}
\includegraphics[width = 80mm]{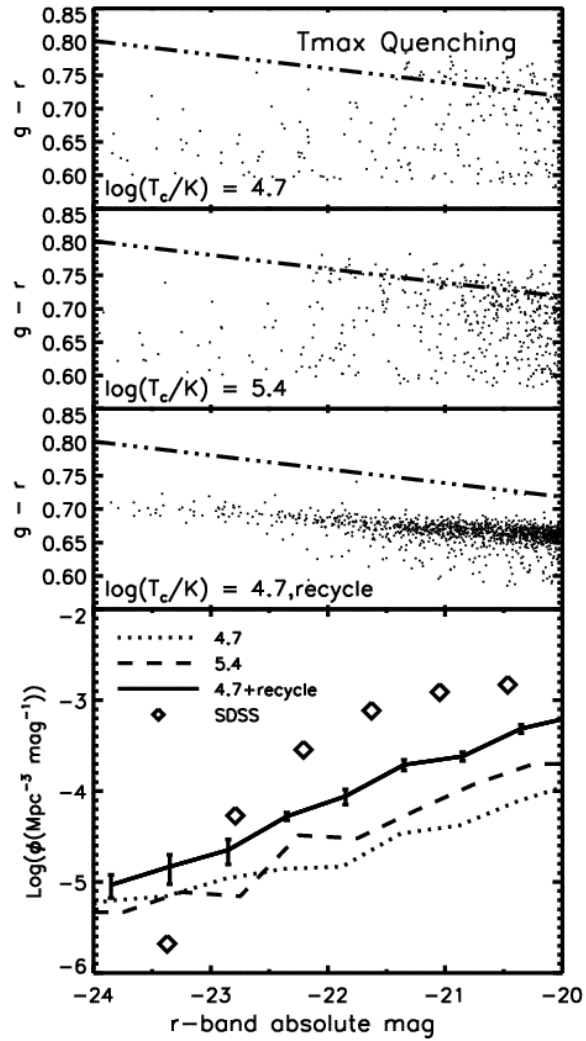}
\caption{$T_{\rmn{max}}$ CMDs and LFs, analagous to Figures
\ref{fig.merger_cmdlf} and \ref{fig.mass_cmdlf}.  We quench star
formation from gas particles that achieved a maximum temperature above
a critical value, $T_c$, for $T_c=10^{4.7}$ and 10$^{5.4}$ K.  These
first models fail to produce a substantial red sequence because at all
masses, galaxies accrete fresh cold gas from the intergalactic medium
and form it into stars.  Although this model may suppress global star
formation, it fails to suppress \emph{all} the star formation in a
given galaxy.  In the third panel we show a model where, along with a
$T_{\rmn{max}}=10^{4.7}$ threshold, we also quench star formation from
any gas particles that were recycled from a galactic wind.  Although
this model shows a reasonable red sequence in the CMD, it strongly
suppresses the formation of blue cloud galaxies (not shown). }
\label{fig.tmax_cmdlf}
\end{figure}

Figure \ref{fig.tmax_cmdlf} shows results of $T_{\rmn{max}}$ (i.e.
hot mode) and wind mode quenching.  The top two panels show CMDs
for two different threshold temperatures, $10^{4.7}$~K and $10^{5.4}$~K.
These are chosen as the bimodality separation in $T_{\rmn{max}}$
in our simulations and the (no wind, no metal cooling) simulations
of \citet{keres09_coldmode}, respectively.  The third panel shows
the extreme case of quenching both hot mode with a $10^{4.7}$~K
threshold, {\it and} quenching all wind mode as well.

Quenching hot mode alone does not produce a red sequence; it is not
bimodal, and the number densities of red galaxies are far too small
at all but the largest masses.  This is because cold mode, either
pristine or in the form of recycled winds, continues to provide
significant accretion at late times.  Hence the idea that simply
keeping gas hot in a hot halo reproduces a red sequence does not
appear to be viable, since it takes only a fairly small amount of
ongoing star formation to make a galaxy blue.

Quenching both hot {\it and} wind mode produces something that looks
like a red sequence, with an amplitude, slope, and scatter that is
(coincidentally) comparable to the merger quenching case.  However,
the red galaxy LF continues to have the wrong shape; it is roughly
just a constant factor higher than in the hot mode-only quenching
case.  \citet{oppenheimer10} anticipate this failure of wind-mode
quenching, finding that \emph{allowing} wind mode accretion leads to the
best-matching stellar mass functions below the turnover mass $M^*$.
\citet{keres09_coldmode} highlighted a numerical problem in GADGET-2
in which cold clumps form owing to thermal instabilities that then
rain down ballistically onto the central galaxy; the trouble is that
the clumps always occur near the resolution limit, even as the
resolution varies substantially.  Hence this cold drizzle may be a
numerical artifact, although \citet{keres09_coldclouds} used very
high-resolution simulations to show that at least some of it is likely
to be real.

Accretion mode quenching appears to be the least promising of our
various quenching mechanisms, so we do not elaborate on these results
any further.  In the discussion we examine some implications of its
failure (\S\ref{sec.failures}).

\subsection{Variants on quenching models} \label{variants}

\begin{figure}
\includegraphics[width=80mm]{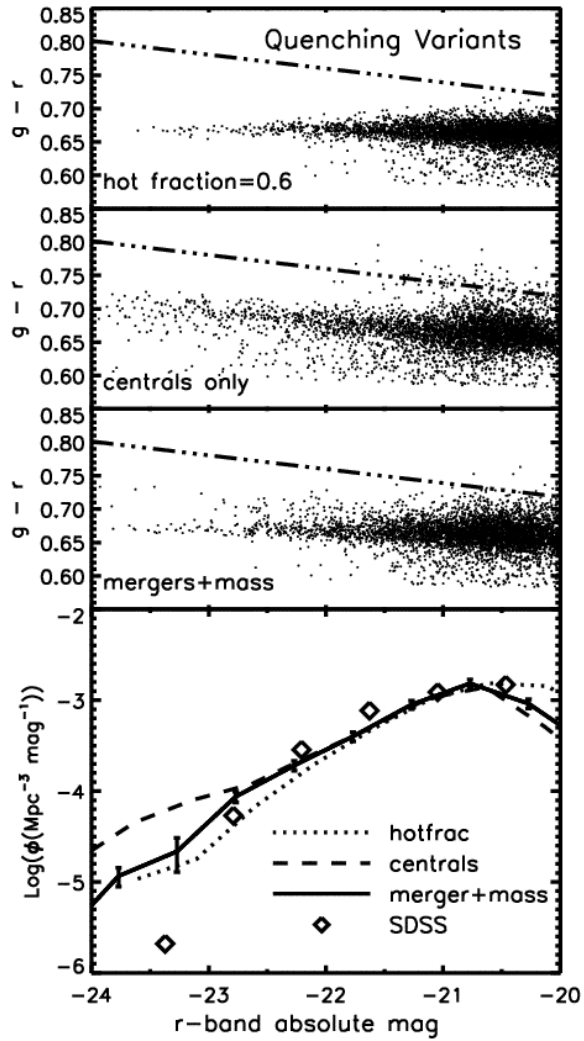}
\caption{ Red Galaxy CMDs and LFs for variants of our quenching models.  The top panel shows the red sequence created by quenching star formation in haloes with a hot gas fraction above 0.6, the second panel shows halo mass quenching where only central galaxies undergo quenching, and the third panel shows a model where both a merger \emph{and} a halo mass above $10^{12} M_{\sun}$ are required to quench star formation.  Although these models show promise at matching the LF at $r>-22$, they share the problems of our simpler models at the bright end: an excess of bright galaxies.
}
\label{fig.variant_cmdlf}
\end{figure}

\begin{figure}
\includegraphics[width=80mm]{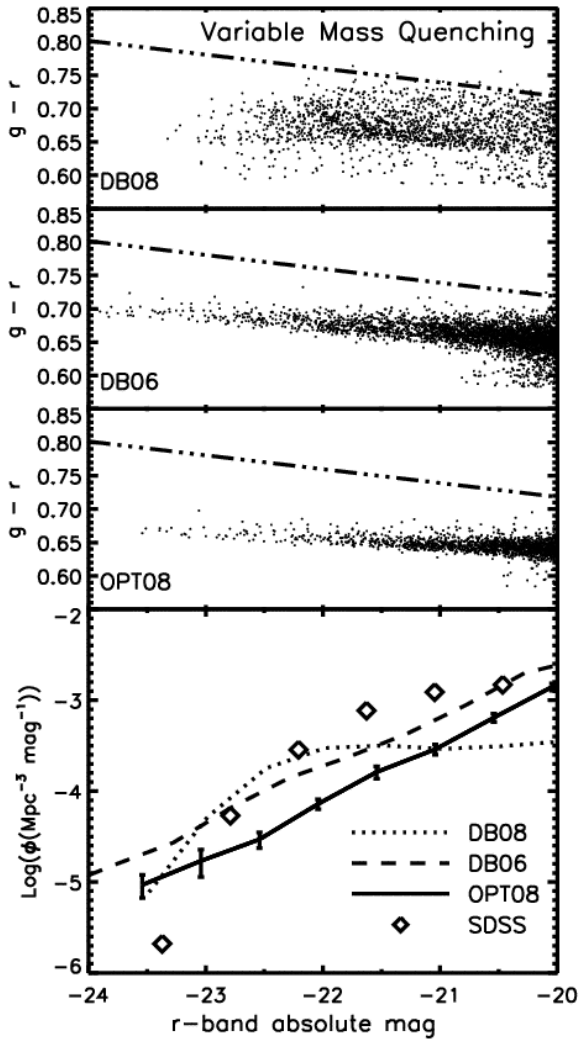}
\caption{Red galaxy CMD and LF for variable mass halo mass quenching.
Instead of a constant critical halo mass above which shocks quench
star formation, the critical mass varies with redshift.  The top panel
uses a model from \protect\citet{dekel08} where the critical mass
required to sustain a hot halo increases with cosmic time.  This model
yields a dearth of galaxies fainter than $r=-22$.  In the other two
models, cold flows are assumed to penetrate hot haloes at
high redshift due to the higher relative densities of cosmic
filaments.  This effectively causes the critical mass above which
quenching occurs to decrease with cosmic time.  We show two different
versions of the critical mass as a function of redshift, based on
\protect\citet{dekel06} and \protect\citet{ocvirk08}, which differ only by the metallicity assumed for cosmic gas.  These models exhibit a nearly-constant slope in the LF.}
\label{fig.mvary_cmdlf}
\end{figure}

After considering our basic models with few free parameters, we attempted
several variations on our quenching models to try to improve agreement
with observations.  In the end there were no obvious successes, though
the results illustrate some interesting trends.

As discussed above, halo mass quenching leads to characteristic clumping
in the red sequence not observed in nature.  This clumping arises from
the sharp critical dark matter halo mass cutoff.  While observational
uncertainties could smear out the clumping, we investigate two more
physically-motivated reasons why the critical mass cutoff might not be
so sharp.

Physically, the halo mass threshold is representative of the mass
scale above which a stable hot halo forms.  Preventative feedback
mechanisms can operate more effectively in the presence of a hot
halo~\citep{keres05,dekel06}.  However, there is non-trivial scatter
between halo mass and hot gas fraction~\citep[e.g.
see][]{keres09_coldmode}.  Hence we tried a model variant in which
we quench galaxies in halos with a hot gas fraction $f_{\rmn{hot}}
= M_{\rmn{hot}} / (M_{\rmn{hot}} + M_{\rmn{cold}})$ above a critical
fraction.  We attempted models with critical fractions 0.2, 0.4,
0.6, and 0.8, which correspond roughly to halo masses
$10^{11.5}$--$10^{12.3} M_{\sun}$ but with a scatter of $\sim
$0.2--0.3 dex.  The results are shown in Figure \ref{fig.variant_cmdlf}
(with the CMD in the top panel), for a critical hot gas fraction
of 0.6.  This variant smooths out the characteristic absolute
magnitude or stellar mass at which blue galaxies move to the red
sequence, yielding a less clumped red sequence.  However, the slope
and scatter (and trends with luminosity) remain essentially unchanged.
Also, the LF still shows the excess of very bright galaxies as noted
in the halo mass quenching case.  Hence besides the aesthetic appeal
of removing the clump in the CMD, it does not fare any better (or
worse) than halo mass quenching.

As another variant, we examined a scenario in which we quench only in
central galaxies of halos, not satellite galaxies.  This is shown in
Figure \ref{fig.variant_cmdlf} (second panel) for a critical mass of
$10^{12} M_{\sun}$.  The CMD in this model shows greater scatter in
the red sequence and a much flatter luminosity function slope than our
preferred model where satellites are quenched as well.  Galaxies are
able to build up more mass before ending up on the red sequence, and
massive central galaxies end up merging with more massive satellites
than in the preferred model.  This failure is disappointing, as most
scenarios of halo mass quenching only quench the central galaxy, and
observations favor a more gradual quenching of
satellites~\citep[e.g.][]{weinmann09}.  Indeed, SAMs that quench
satellite galaxies in halos have difficulties reproducing the colour
distributions of satellites~\citep{weinmann06,baldry06}, which is a
drawback that is shared by our original halo mass quenching scenario.
We do note that the red sequence slope seems to be in better agreement
with data, as larger galaxies acquire larger (unquenched) satellites
that are more metal-rich.

We also investigated a hybrid model that requires both a major
merger and a critical halo mass for quenching.  That is, we quench
only if a major merger occurs inside a halo with mass $>10^{12}
M_\odot$ (here we assume $f_{\rmn{gas}}=1$).  Physically, this model
mimics the action of a hot halo to prevent post-merger gas accretion
from the IGM, perhaps by continued energy injection from an AGN.
In our original quenching model, such post-merger accretion is
always prevented, but here we only prevent accretion in larger halos
that can support a virial shock.  This model is perhaps closest to
what is envisioned in current ideas for quasar plus radio mode
quenching, in which a merger initially transforms the galaxy into
an elliptical and quenches star formation, but only when the resulting
halo is sufficiently large to form a virial shock does the resulting
halo gas stay hot e.g. via low-level AGN activity.

Figure~\ref{fig.variant_cmdlf}, third panel, shows the results of this
model.  The results are fairly similar to halo mass quenching, showing
that it is this aspect that is the limiting factor for quenching;
mergers are frequent enough that it does not add a stringent
criterion.  The red sequence shows slightly more scatter than in halo
mass quenching, and there is no evidence for clustering at a
particular $r$-band magnitude; the slope is still incorrect, and (as
always) the blueness problem persists.  The LF is also similar to the
halo mass quenching case, with an excess of very bright galaxies and a
possible dearth of $\sim L^*$ systems.  This model may, in fact, be
the best-fitting model as well as the most physically-motivated, but
the results are qualitatively very similar to those of our simpler
models.

Finally, we consider three variants of halo mass quenching where
the critical halo mass varies with redshift.  The first is motivated
by \citet{dekel08}, who argue that the critical halo mass for
quenching owing to gravitational clump heating varies slightly with
redshift due to the evolving cosmic density of gas (see their
Figure~3).  In this scenario, the quenching mass varies from $M_c
\simeq 10^{11.8}M_{\sun}$ at $z=3$ to $M_c \simeq 10^{12.8} M_{\sun}$
at $z=0$.  For $z>3$ we use the lower value, $M_c = 10^{11.8}
M_{\sun}$.  Figure \ref{fig.mvary_cmdlf}, top panel, shows the
results.  The shape of the LF is affected most strongly, as there
is now a significant dearth of red galaxies fainter than $r\simeq
-22$.  The high $M_c$ at late times makes the fainter end of the
red sequence similar to the fixed $M_c=10^{12.5} M_\odot$ case,
i.e., it does not populate the red sequence enough.  We do note
that it produces a larger red sequence slope that is in better
agreement with data, although the scatter is larger.  Overall, this
model does not fare as well as our favored halo mass quenching model
with a constant $M_c=10^{12} M_\odot$, although it does illustrate
that varying the quenching mass with epoch can produce notable
changes in the red sequence (at the aesthetic expense of introducing
more parameters).

In another paper, \citet{dekel09_nature} argue that $M_c$ evolves with
redshift in the opposite sense, namely that $M_c \approx 10^{11.7}
M_{\sun}$ out to $z=1.5$, but at higher redshifts it increases rapidly
owing to the ability of cold streams to penetrate through hot halos at
early epochs (see their Figure~5, and also \citealt{dekel06}). From an
eyeball estimate of their conjectured hot halo mass limit in the
presence of cold streams, we take $M_c = 10^{11.7} M_{\sun}$ for
$z<1.5$, and $\log M_c \simeq 1.2z+9.9$ for $z>1.5$.  We also show a
variation from \citet{ocvirk08} where gas in cosmic filaments in a
hydrodynamic simulation is found to have a lower metallicity than
\citet{dekel06} assume.  In this formulation, $M_c \simeq 10^{11.5}$
for $z<3$, and $\log M_c \simeq 1.4z + 7.3$ for $z>3$.  The lower
panels of Figure \ref{fig.mvary_cmdlf} show the results in these
cases: A red sequence is produced, but the CMDs have a very small
colour scatter, and more dramatically the LFs are power laws!  The LFs
of the two parametrizations differ primarily in their normalization.
These models exacerbate the problem of producing a knee in the LF over
constant-$M_c$ models.

None of these quenching model variants stands out as obviously
superior to the others.  Since the simple models perform
as well as the variants, and because they involve simpler prescriptions,
we use merger quenching with $f_{gas}=1$ and $r=3$ and halo mass
quenching with $M_c=10^{12} M_{\sun}$ as our preferred models.  For
the remainder of the paper, we focus on these two models.


\subsection{The Blue Cloud} \label{sec.dust}

\begin{figure*}
\begin{minipage}{7in}
\centering
\includegraphics[width = 5.75in]{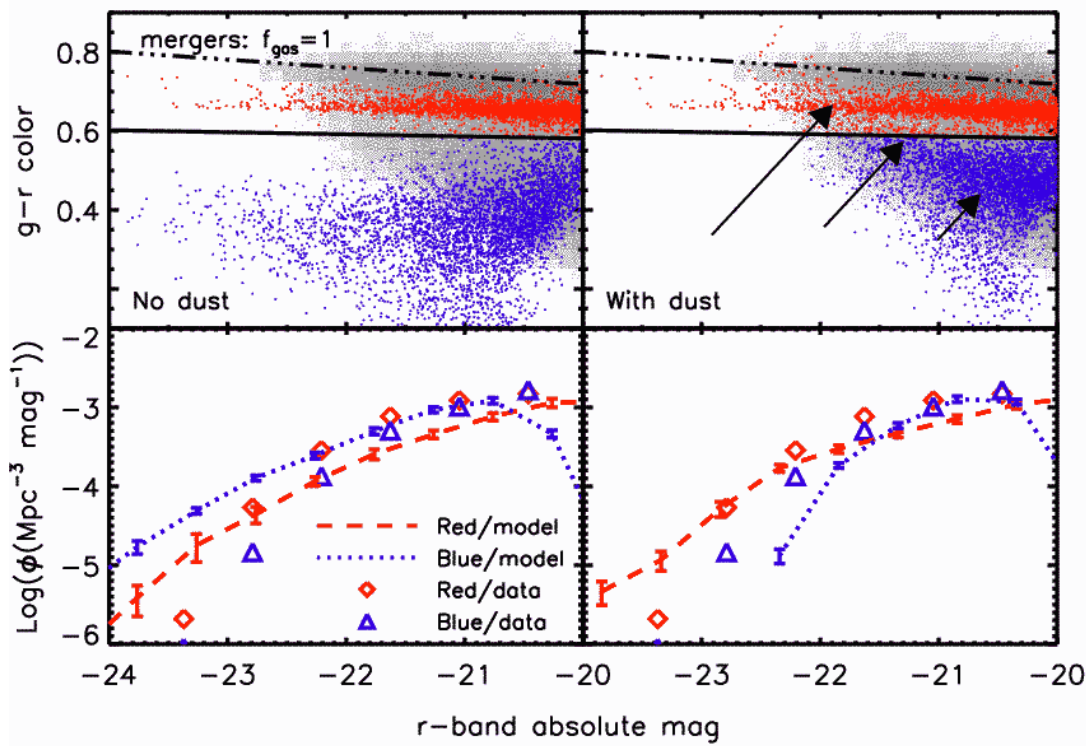}
\caption{ Dust comparison of CMDs and LFs for merger quenching with
$f_{\rmn{gas}}=1$.  Our dust model (right panels) shifts bright blue
galaxies up and to the right in the CMDs, introducing contaminants to
the bright end of the red sequence.  Arrows illustrate the change in
position (from tail to head) due to dust for three galaxies, with
star-formation rates $\approx$3, 9, and 15 $M_{\sun}$ yr$^{-1}$ (right to
left).  The luminosity functions reflect the increase in bright red
galaxies due to contamination, as well as the suppression of bright blue ones.}
\label{fig.dust_comparison_merger}
\end{minipage}
\end{figure*}

\begin{figure*}
\begin{minipage}{7in}
\centering
\includegraphics[width = 5.75in]{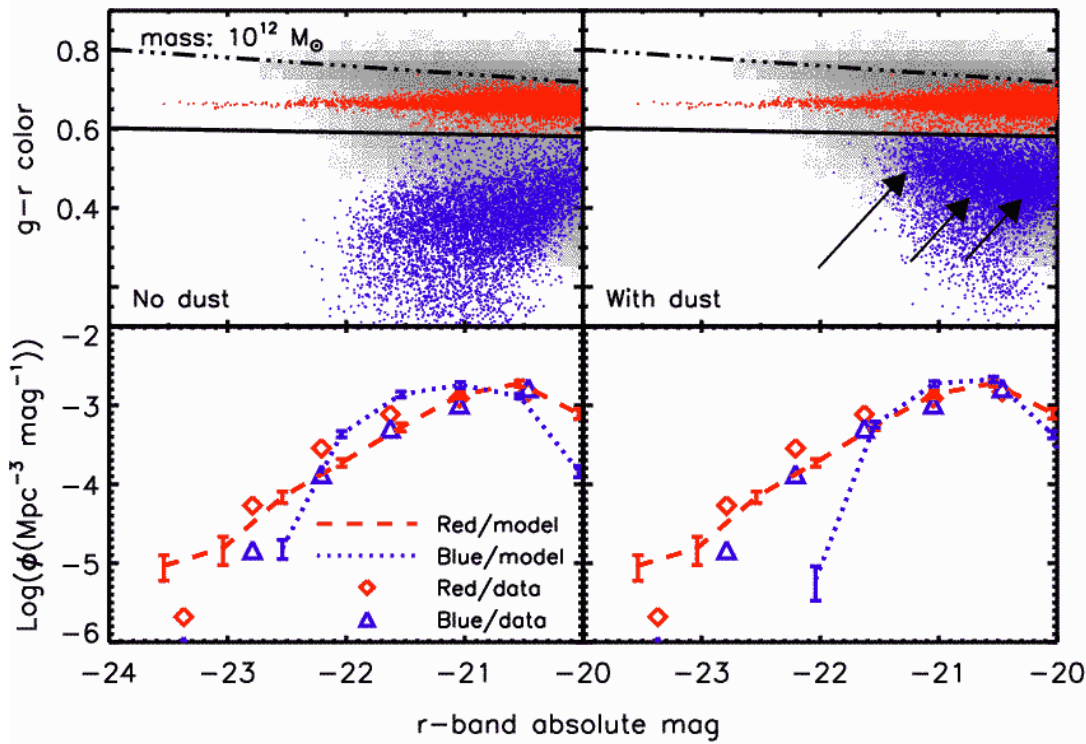}
\caption{ Dust comparison of CMDs and LFs for halo mass quenching
with $M_c=10^{12} M_{\sun}$.  Arrows show the change in position for
three galaxies, with star-formation rates $\approx$2, 4, and 7.5
$M_{\sun}$ yr$^{-1}$ (right to left).  Even without dust (left
panels), this model does not produce exceptionally bright blue
galaxies: once blue galaxies attain a critical stellar mass
corresponding to $M_c$, they move to the red sequence.  Our dust model
(right panels) suppresses the bright blue galaxies, leading to a steep
cutoff in the blue galaxy luminosity function.}
\label{fig.dust_comparison_mass}
\end{minipage}
\end{figure*}

So far we have focused on the red sequence.  Our merger and halo
mass quenching models also produce a star-forming blue cloud of
galaxies that is separated from the red sequence, as shown in the
left panels of Figures~\ref{fig.dust_comparison_merger} and
\ref{fig.dust_comparison_mass}.  Here we examine the properties of
the blue cloud in these scenarios.

Computing the observed luminosities of blue cloud galaxies is
complicated by uncertainties in dust extinction.  Our simulation
tracks metallicity, which is correlated with dust extinction.  
There are also empirical correlations between UV or blue-band luminosity
and extinction.  Since in general the bright blue galaxies are quite
metal-rich, the exact form and nature of the assumed extinction makes
a significant difference in their resulting brightness and colours.
We only apply an extinction correction to blue galaxies, leaving
quenched galaxies unaffected.  Also, we employ \citet{calzetti00}
reddening law where the optical depth due to dust varies as wavelength
$\lambda^{-0.7}$.

We first tried employing a correlation between metallicity and
extinction from SDSS, including scatter, as described in
\citet{finlator06}.  However, we found that the resulting blue cloud
had a very large scatter, and a substantial number of previously
blue galaxies ended up redder than the red sequence.  It is possible
that second parameter correlations exist in metallicity versus
extinction that are not accounted for, but we did not explore this
further.

Our preferred method, i.e. one that produced a blue cloud similar
to observed, uses an empirical correlation between UV luminosity
and dust extinction given by \citet{wang96}, also described in
\citet{finlator06}.  This prescription moves bright blue galaxies
more than dimmer ones, so that the bluest galaxies are not necessarily
the brightest ones.  The resulting overall shape of the blue sequence
matches the observed SDSS blue sequence reasonably well.

Figure \ref{fig.dust_comparison_merger} compares CMDs and LFs for our
merger quenching model with and without dust.  Without dust, the brightest
blue galaxies in the universe would have $r$-band luminosities brighter
than the brightest red galaxies, and they would be $\sim 0.3$ magnitudes
bluer than the red sequence in $g-r$.  With dust applied, the brightest
star-forming galaxies are scattered into the red sequence, making up
$\sim$6\% of red sequence galaxies; this is reasonably consistent with
observational estimates \citep[cf.][]{bell04_dusty, brammer09}.  Furthermore, the dust obscuration ensures
that massive blue galaxies are not brighter than the red sequence.
The luminosity functions reflect this difference markedly.  In the no-dust
case, blue galaxies dominate over the bright end of the red sequence, but
the presence of dust shifts the bright blue objects to lower luminosities.

Figure \ref{fig.dust_comparison_mass} compares the dust and no-dust
cases for halo mass quenching.  Even the no-dust case lacks bright
blue galaxies like those seen in the merger model and in observations,
and once dust is included the problem becomes significantly more severe.
As discussed in \S\ref{sec.results.mass}, once galaxies attain a stellar
mass corresponding to the critical halo mass, they move out of the blue
cloud and onto the red sequence.  Application of the dust prescription
pushes blue cloud galaxies up and right in the figure, creating a diagonal
envelope.  Since dust tends to move the brightest galaxies the most, the
dearth of blue galaxies intrinsically brighter than $r \sim -22$ leads
to fewer intrinsically blue interlopers on the red sequence ($<2$\%).

As an aside, we note an interesting study by \citet{maller09} of galaxy
orientation in relation to dust obscuration, and its impact on derived
galaxy properties.  Notably, they find that structural parameters like
axis ratio and S\'{e}rsic index significantly impact measured colours and
magnitudes, and they find an average of 0.2-0.3 magnitudes of extinction
in the SDSS $g$ and $r$ bands.  This study predicts considerably more
bright blue galaxies than are generally inferred from studies that
do not take into account orientation-dependent effects.  We do not
account for such effects in our work, as we use a simple dust screen
model to account for obscuration.

While such comparisons are illustrative, the sensitivity to the exact
extinction prescription used makes robust interpretation of the
discrepancies difficult.  As an alternative approach, we can perform
the comparison versus stellar mass functions rather than luminosity
functions.  The stellar masses derived by
\citet{kauffmann03_sf_stellarmass} implicitly account for dust
extinction on a galaxy-by-galaxy basis, albeit with some assumptions
about the reddening law.  \citet{baldry08} show that these masses,
determined by comparing star formation history templates based on the
BC03 models to absorption features in the galaxy spectra, yield
similar mass functions to other mass derivations using photometry,
different population synthesis models, or different spectral features.



\begin{figure}
\includegraphics[width=84mm]{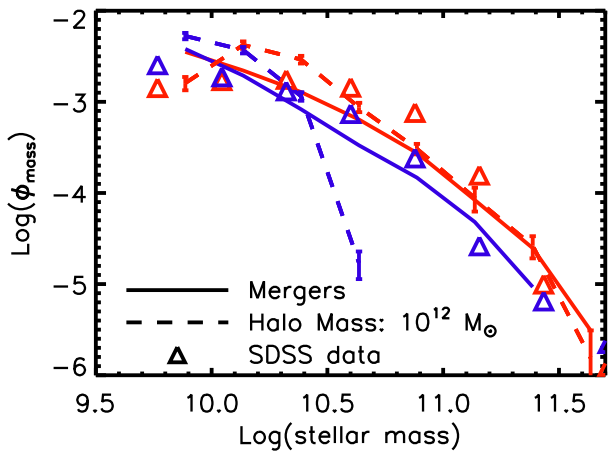}
\caption{ Stellar mass functions for merger quenching (solid lines),
halo mass quenching (dashed lines), and SDSS data (symbols).  We
split the galaxy populations into blue and red, as indicated by the
colour of the lines.  For merger quenching, the shape of the blue
galaxy mass function tracks that of red galaxies, roughly in agreement
with observations.  Halo mass quenching, however, yields a sharp
cutoff in the blue galaxy mass function.}
\label{fig.massfunc}
\end{figure}

In Figure \ref{fig.massfunc} we show stellar mass functions (the
number of galaxies per Mpc$^3$ per logarithmic stellar mass bin),
separated into red and blue galaxies, for merger quenching, halo
mass quenching, and SDSS galaxies.  When dividing our simulated
galaxy sample into red and blue, we do not use our dust prescription,
so that Figure \ref{fig.massfunc} shows the mass functions for
intrinsically blue and red galaxies.  Given the small level of
contamination of actively star-forming galaxies on the red sequence,
this choice does not significantly impact the results.

In our models, the red galaxy mass functions behave similarly to
the luminosity functions, as expected since for these galaxies
$r$-band absolute magnitude is a good tracer of stellar mass.  The
red mass functions do not have a sharp knee, and the slope for halo
mass quenching is slightly steeper than that for merger quenching.
In fact, above $M^*$ (the characteristic stellar mass in a Schechter
function fit, $\approx 10^{10.7}M_\odot$), the red galaxy mass
functions in the two models are remarkably similar.

The blue galaxy mass functions reveal a crucial difference between
our halo mass and merger quenching models.  Halo mass quenching
produces a precipitous drop in the blue mass function above $\sim
10^{10.5} M_{\sun}$, a critical stellar mass corresponding to the
critical halo mass $10^{12} M_{\sun}$.  Mergers, in contrast, yield
a blue stellar mass function whose shape traces that of the red
stellar mass function (but with slightly lower normalization) for
massive galaxies.  The merger scenario markedly better reproduces
the observed blue galaxy mass function.  At $M_* > 10^{11} M_{\sun}$
there are fewer blue galaxies, but \emph{not} zero blue galaxies.

The overall conclusion from the stellar mass function comparison is
the same as obtained from the luminosity function comparison: the
merger quenching model produces red and blue stellar mass functions
that broadly agree with data, but the halo mass quenching produces a
sharp truncation in the large blue galaxy population that is in
disagreement with data.  We emphasize that this truncation persists
even in our variant quenching models which smear out the clumping
associated with a halo mass cutoff (the hot fraction variant).  This
generic result is difficult to avoid given a relatively tight
correlation between halo mass and stellar mass seen in the
simulations.  Unless this correlation has much greater scatter in
nature than in our models, a simple halo mass quenching model will
have difficulty matching these data.

Of course, the strict halo mass cutoff we adopt here
represents an oversimplification.  Even when we tie quenching to the
physically more relevant fraction of hot gas, we truncate star
formation instantaneously and forever.  Real galaxies are unlikely to
behave this way.  After a stable hot gas halo has formed and quenched
the fuel supply, star formation may continue with an existing cold gas
reservoir.  Furthermore, passive galaxies may undergo mergers with
gas-rich satellites, momentarily reinvigorating star formation and
adding young, blue stars.

Introducing a gigayear delay for the onset of quenching after
a halo reaches the mass threshold, we still find no blue galaxies
above $10^{10.7} M_{\sun}$ (while allowing more red galaxies to grow
stellar masses above $10^{12} M_{\sun}$).  Such a time delay may not
be much more realistic than our base model, but simulated galaxies
around $10^{10.5} M_{\sun}$ typically have gas fractions well below
0.5, so that they cannot grow by the required factors $\sim 10$ using
just their existing reservoirs of cold gas.  This suggests that the
steep cutoff in the blue galaxy mass function is robust.  For pure
halo mass quenching to be viable, we require a mechanism that chokes
off the fuel supply of most of the massive galaxies in hot haloes, but not
all.  A more detailed treatment of massive, star-forming galaxies
awaits fully hydrodynamic simulations that incorporate quenching
mechanisms.

\section{Discussion}
\label{sec.discussion}

\subsection{Origins of the red sequence}
\label{sec.origins}

\begin{figure*}
\begin{minipage}{7in}
\centering
\includegraphics[width=5in]{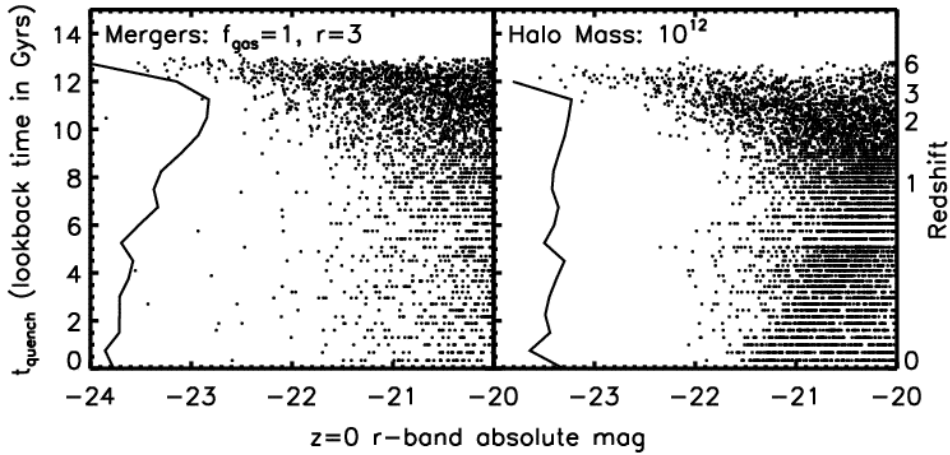}
\caption{ Lookback time of initial quenching vs. $z=0$ $r$-band absolute
magnitude for merger quenching and halo mass quenching.  For
mergers, $t_{\rmn{quench}}$ is the lookback time of the first major
merger.  For halo mass quenching, $t_{\rmn{quench}}$ is the
lookback time when the galaxy's dark matter halo first exceeded the
critical halo mass of $10^{12} M_{\sun}$.  Along the $y$-axis, we show
histograms of $t_{\rmn{quench}}$.  The brightest galaxies all entered
the red sequence at early times ($z>2$).  Merger quenching yields a
peak quenching era at $z\approx 3$, whereas a halo mass limit quenches
galaxies with a rate nearly constant in time.}
\label{fig.tquench}
\end{minipage}
\end{figure*}

\begin{figure*}
\begin{minipage}{7in}
\centering
\includegraphics[width = 6in]{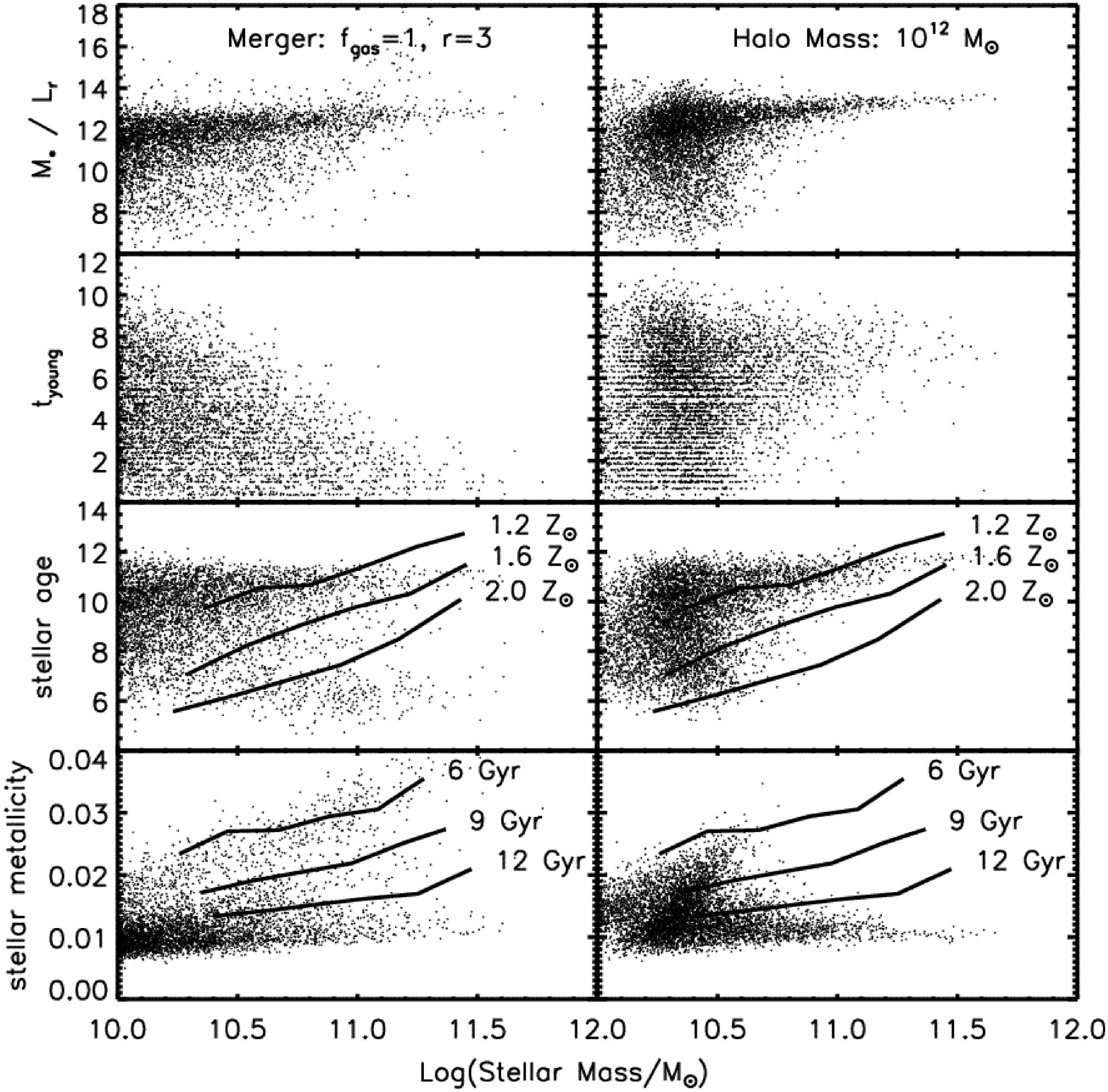}
\caption{Redshift zero $r$-band mass-to-light ratios, age of the
youngest star particle, mean stellar age, and mean stellar metallicity
vs. stellar mass for galaxies in the merger quenching (left) and halo
mass quenching (right) models.  All panels include only red sequence
galaxies.  In the mean stellar age panels, lines represent the mean
stellar age required to get the correct colours of the SDSS red
sequence assuming a fixed metallicity along the red sequence of 1.2,
1.6, or 2.0 $Z_{\sun}$ (as labeled on the plot; see text), where
$Z_{\sun}\approx 0.012$.  The metallicity plot shows analogous lines
assuming a constant age along the red sequence of 6, 9, or 12 Gyrs.}
\label{fig.stuff_vs_mstar}
\end{minipage}
\end{figure*}

Observations now suggest red and dead galaxies have existed since
at least $z\simeq 2$ \citep{kriek08, williams09, brammer09}, and
the formation time for their stellar populations is quite old.  When
does the quenching take place in our models?

Figure \ref{fig.tquench} shows the lookback time at which initial
quenching occurred for each of the quenched galaxies in our two
preferred quenching models.  For the brightest red galaxies, quenching
occurred at $z\geq$ 3 in both models.  In halo mass quenching,
essentially zero galaxies brighter than $r=-22$ were quenched after
$z=1.5$, whereas in merger quenching a small number of bright
galaxies came to the red sequence at late times.  In halo mass
quenching, red galaxies can only achieve the highest masses via
mergers \emph{after} moving to the red sequence because the
effective critical stellar mass ($\sim 10^{10.5} M_{\sun}$) is much smaller than the biggest
galaxies.  In merger quenching, on the other hand, blue galaxies
can attain quite high stellar masses and then move onto the red
sequence as an already-bright galaxy at late times.

Along the $y$-axis we show vertical histograms in $t_{\rmn{quench}}$,
indicating the quenching rate in terms of the number of newly
quenched galaxies as a function of cosmic time.  The merger quenching
rate peaks at $z\sim3$, whereas a halo mass threshold quenches at
a roughly constant rate over time.  Here we caution that the peak
in the merger quenching rate is sensitive to our minimum mass for
merger quenching, which is set by the simulation resolution.  A
higher resolution simulation would likely move the peak to higher
redshift, since small galaxies could quench earlier.  In any case,
the peak will remain as a distinguishing feature between merger and
halo mass quenching.

For a deeper physical understanding of our model galaxies, we turn
to Figure \ref{fig.stuff_vs_mstar}, where we plot $r$-band stellar
mass-to-light ratios, $t_{\rmn{young}}$ (the age of the youngest
star particle), mean stellar age, and mean stellar metallicity vs.
stellar mass for our galaxies in our two preferred quenching models.
We estimate mass-to-light ratios in solar units by dividing stellar
mass by $L_r$, which is given by $\log(L_r / d\nu) = (r-C)/(-2.5)$.
Here $r$ is the galaxy's $r$-band absolute magnitude, $C=-48.6$ for
AB magnitudes~\citep{oke74}, and $d\nu = d\lambda (c/\lambda^2)$
is the approximate frequency width of the SDSS $r$-band filter.  We
take $\lambda = 6250$ and $d\lambda=1500$ angstroms for the $r$-band.
In both quenching models, we find a mass-to-light ratio that increases
slowly with stellar mass, and is fairly tight above $10^{10.5}
M_{\sun}$ where most galaxies have been quenched for some time.
Halo mass quenching displays a clump of M/L just below the quenching
(stellar) mass, since these galaxies have recently undergone
quenching.

In the second row, we highlight a distinction in the age of the
youngest stars between the merger quenching and halo mass quenching
models.  Massive galaxies in the halo mass quenching model lack any
population of young stars: all the galaxies they accrete have old
stellar populations, since even satellites that will be accreted
later are quenched in massive halos at an early time.  Galaxies
quenched via mergers, however, include trace populations of young
stars.  Even though mergers quenched the star formation in these
massive galaxies at $z>2$, they obtain young stars via accretion
of younger satellite galaxies.

In the bottom panels of the figure, we examine ages and metallicities
to understand why our models do not match the slope and normalization
of the real red sequence.  We include tracks showing the mean ages
and stellar metallicities vs. stellar mass required to reproduce
the observed red sequence.  Specifically, on the plot of mean stellar
age vs. stellar mass, we assume a uniform metallicity for all red
sequence galaxies, and ask:  what mean ages would those galaxies
need to have to get the correct red sequence colours using the BC03
models?  To answer this question, we create a grid of artificial
single stellar populations (SSPs) with a variety of masses and
randomly chosen metallicites and ages.  This grid densely samples
the region of the CMD where real red sequence galaxies lie.  We
then take our fit to the SDSS red sequence from Figure
\ref{fig.red_seq_scatter}, and for each absolute magnitude bin we
identify all the artificial SSPs which fall within the scatter of
the median in $g-r$ and within 0.003 of the assumed metallicity.
We compute the mean stellar mass and mean stellar age of these
SSPs, and plot as a connected line in the figure.  The plotted
values have a scatter of $\sim$1 Gyr.  We follow an analogous
procedure for the tracks in the metallicity vs. stellar mass plots.

These tracks tell us what slope and normalization in age (metallicity)
would be required to get the correct red sequence slope, \emph{assuming}
the red sequence has a constant metallicity (age).  Of course, in
real galaxies both age and metallicity may vary with stellar mass,
but this is intended to illustrate the general trends.  If age
gradients along the red sequence are small, then we require a factor
$\sim$2 difference in metallicity from the brightest galaxies to
those an order of magnitude less massive in order to reproduce the correct
slope.  If metallicities are roughly constant across the red sequence,
then we require variation of 3--4 Gyrs between the bright and faint
galaxies for typical galaxy ages. Finally, if a typical age is
10~Gyr, then the simulation metallicities would need to be increased
by $\sim\times 2$ in order to match the observed red sequence 
amplitude.  We will discuss the implications of these trends
in \S\ref{sec.failures}.

\subsection{The blueness problem} \label{sec.blueness}

In all cases our models yield red sequences too blue by about 0.1
magnitudes in $g-r$, an error of $\sim$10\% in the luminosity ratio
between those bands.  For these galaxies, colours are determined
by matching each constituent star particle's metallicity and age
to grids from BC03 stellar population synthesis models.  While the
discrepancy might originate with uncertainties in the population
synthesis models \citep[e.g.][]{charlot96, conroy09a}, we attempted
to mitigate this possibility by using the same models as
\citet{blanton05_vagc} for the VAGC.  We also tried the updated Charlot \&
Bruzual models (graciously provided to us by S. Charlot)
that employ an updated treatement of thermally pulsating (TP) AGB
stars, but these showed very minor difference compared to BC03 for
our red sequence galaxies since TP-AGB stars do not contribute
significantly to the luminosity in SDSS bands.  Hence we explored
other possible reasons for the blueness problem.

There are two ways to make a galaxy too blue: give it a younger
age, or a lower metallicity.  Perhaps our simulations simply have
distributions of galaxy ages and metallicies that are wrong.  As
shown in Figure \ref{fig.stuff_vs_mstar}, our brightest red galaxies
have mean stellar ages $>$ 10 Gyrs.  This leaves little room to
maneuver, as galaxies cannot get much older given the age of the
Universe.

The story is different for metallicity.  The tracks in Figure
\ref{fig.stuff_vs_mstar} suggest that the metallicities of our
galaxies are systematically low.  Very few galaxies show metallicities
above 0.02 (metal mass fraction), but BC03 models require that the
most massive galaxies have metallicities significantly above this
to explain their red colours.  Why might these galaxies be unrealistically
metal poor?

As described in \citet{finlator08}, metallicities in un-quenched
galaxies are determined by an equilibrium between less enriched gas
accreted from the IGM and gas enriched by ongoing star formation.
Metallicities tend to rise as a galaxy grows in mass, and even at
fixed mass galaxies at lower redshift are more metal rich~\citep{dave07}.
Once quenching occurs in our models, the metallicities are frozen
into the existing star particles.  If our simulations quenched
galaxies too early, the galaxies would not have built up their
metallicities sufficiently.

Another uncertainty is the adopted supernova metal yields.  We use
a set of yields from various authors, most notably \citet{chieffi04}
yields for Type~II supernovae that dominate the stellar metal budget
of these galaxies.  The yields are not all that well constrained,
and it is possible that they are higher than we have assumed.  We
note that our simulations assume \citet{asplund05} abundances, for
which solar metallicity is a metal mass fraction of 0.0126, as
opposed to the 0.02 fraction that is assumed for solar metallicity
in the BC03 models.  We assume that the metallicity scale in the
BC03 models represents an absolute scale (S. Charlot \& A. Bressan,
private communication), so that e.g. a simulated galaxy that has
``solar" metallicity of 0.0126 is computed using the $0.6Z_\odot$
BC03 templates.

Our enrichment and galactic wind models further complicate the
interpretation of the metal content of simulated stellar populations.
In our simulations, star formation induces both metal enrichment
and galactic winds.  The winds suppress enrichment of subsequently-forming
stars by removing metals from the ISM; this process is poorly
contrained observationally, and our model could be overly efficient
at metal removal.  We note, however, that the ejected metals in our
outflow model match observations of IGM enrichment \citep{oppenheimer08,
oppenheimer09_OVI}, so it is not clear that one can lower the ejected
metals substantially.  Since, however, most metals end up in the IGM
particularly at early times~\citep{opp09_metal_absorption}, it would
not take large changes to the outflow model in order to double the
metallicity retained in galaxies.

Our simulations do reproduce the stellar mass--gas phase metallicity
relation for galaxies at $z\approx2$ \citep{finlator08}, suggesting
that metallicities for star-forming galaxies in our simulation are
generally correct.  However, systematic uncertainties at the 0.2--0.3
dex level could impact the \citet{finlator08} comparison with data
from \citet{erb06}.  These uncertainties emerge from the calibration
of metallicities derived from observations \citep[cf.][]{kewley08},
and because \citet{finlator08} compared all simulated galaxies to
UV-selected observations, which may have higher specific star
formation rates and lower metallicities \citep{ellison08}.  Hence
there is plausibly room for $\sim\times 2$ variations in the
metallicities of galaxies.

Finally, the under-enrichment could be related to the star formation
history in the simulations.  In particular, at least for the merger
quenching, one expects that a starburst may be associated with the
merger event~\citep[e.g.][]{mihos96}.  We do not explicitly include
such a starburst when we apply our quenching, and it could consume a
significant amount of gas to create an enhanced metallicity.  However,
we note that we only quench galaxies typically well {\it after} the
merger event, since we quench only at fixed time intervals of the
simulation snapshots.  While our simulations lack the resolution to
track the detailed star formation history during the merger, the {\it
integral} of the SFH is typically close to what would be obtained at
much higher resolution, since it is primarily limited by the overall
gas supply.  Hence while the resolution limits may modestly impact
star-formation histories, it is unlikely that higher resolution can
boost the metal production (or stellar production) by a factor of two.
Indeed, a simple comparison of our base simulation with
another simulation with better mass resolution by a factor $>3$ shows
negligible differences in the mass-metallicity relation for
star-forming galaxies at $z=1$ \citep[see also][]{dave07}.  For resolved galaxies at fixed
stellar mass, fractional differences in metal content between the two
simulations are no larger than about 5 per cent, considerably smaller
than the mean fractional scatter within a single simulation of 13 per
cent.

It is possible that there are some effects in the observations that
make the observed galaxies too red.  Deriving reliable absolute
magnitudes from apparent magnitudes requires several steps and
important assumptions.  One must use some galaxy template to calculate
a galaxy's redshift/distance and k-correction \citep{hogg02,
blanton07} from photometric data.  In the VAGC, this is done by
creating a suite of star formation histories with corresponding
spectra from BC03 models, running a principle components analysis
with the data to determine which star formation histories are most
representative, then fitting each galaxy's photometric data to the
principle components as templates.  The complexity of this procedure
makes it difficult to independently determine the reliability of
the fits or whether the chosen templates are sufficiently representative.
\citet{blanton05_vagc} do estimate uncertainties in reported absolute
magnitudes for each galaxy in the VAGC; for $g$ and $r$, these are
typically $\sim 0.04$ magnitudes.  Only systematic uncertainties
much larger than these could give rise to the blueness problem,
which seem unlikely.

In summary, the blueness problem is likely to reflect some issue
with the chemical enrichment level of simulated galaxies.  The most
obvious explanation is that the yields are somewhat too low, or
that our outflow model is too efficient at metal removal.  In either
case, while the overall metallicities are too low, the relative
metallicities are basically correct, which means that predictions
for the slope and scatter of the red sequence are robust.  However,
it cannot be ruled out that this is reflective of a more subtle
issue with the star formation or chemical enrichment histories of
our simulated galaxies.


\subsection{Star formation histories} \label{sec.sfh}

\begin{figure}
\includegraphics[width=84mm]{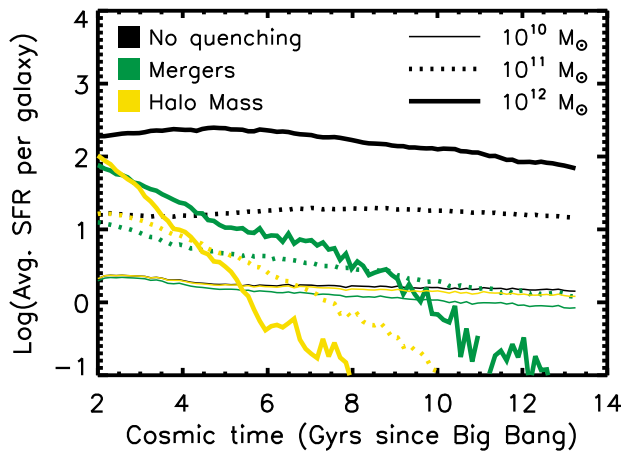}
\caption{Average star formation rates (in $M_{\sun}/$yr) vs. cosmic
time for galaxies in 3 mass bins of width 0.1 dex: $10^{10}, 10^{11},$
and $10^{12} M_{\sun}$.  We sort galaxies into bins by their $z=0$ stellar mass
\emph{in the no-quenching simulation}, then track the mass growth of
those same galaxies through cosmic time in the no-quenching (black),
merger quenching (green), and halo mass quenching (yellow) cases.
By construction, star formation within small galaxies that later merge
to form a large galaxy is counted toward the average SFR.  Both
quenching models strongly suppress star formation in the two massive
bins, but only slightly change the low-mass bin.  Qualitatively, this
behavior mimics the observed ``downsizing'' of star formation.  Halo mass quenching suppresses growth of the most massive galaxies more
strongly than merger quenching because in halo mass quenching a massive galaxy's
satellites will occupy the same halo, and thus be quenched, well before they merge. }
\label{fig.sfr_vs_time}
\end{figure}

By shutting off star formation in the largest galaxies, quenching
mechanisms strikingly impact global star formation.  In our
simulations without quenching, the most intense star formation occurs
in the most massive galaxies, accounting for significant fractions of
the of global star-formation rate density.

We show the effects of merger quenching and halo mass quenching on the
star formation histories of galaxies of different masses in Figure
\ref{fig.sfr_vs_time}.  We first select galaxies from three stellar
mass bins, $10^{10}, 10^{11},$ and $10^{12} M_{\sun}$, at $z=0$ in the
simulation with no quenching model applied.  The bins have a width of
0.1 dex in stellar mass.  We then find the time each star in these
galaxies was formed, translating to star-formation rate (in
$M_{\sun}/$yr) by averaging over intervals of $\sim$100 Myrs.
Dividing by the number of galaxies in each mass bin, we get the
average SFR per galaxy as a function of cosmic time.  This methodology
includes in the SFR any star-formation which occurred in small
galaxies that later merged to form a massive galaxy that falls within one of
our mass bins.

For the quenching models, we track the same galaxies selected above
from the no-quenching simulation, but we ignore the contribution of
quenched stars particles to the SFR.  We thus show the effects of
quenching on specific collections of galaxies, rather than galaxies in
given mass bins selected for each model (the quenching models do
not produce $10^{12} M_{\sun}$ galaxies).

Both merger quenching and halo mass quenching strongly suppress
star formation in massive galaxies, while barely changing the average
SFR in the lowest mass bin.  This effect naturally leads to
``downsizing'' with cosmic time \citep{cowie96, heavens04, juneau05},
where more star formation occurs in lower-mass galaxies at later
times.  Halo mass quenching suppresses SFRs more efficiently than
mergers because it quenches satellite galaxies which later merge
with the central galaxy.  In contrast, merger quenching allows
satellites to form stars until the final merger.  Hence in principle,
deriving accurate mean star formation histories from stacked galaxy
samples in various mass bins can provide strong constraints on
quenching mechanisms.

\subsection{Successes and failures of quenching models} \label{sec.failures}

The most basic result of this paper is that both merger quenching and
halo mass quenching successfully recover qualitative aspects of
observed local CMDs and LFs.  These quenching models greatly improve
upon the results of hydrodynamic simulations that do not have any explicit
quenching, which grossly overproduce the abundance of massive blue
galaxies and fail to produce hardly any massive red galaxies.  At
absolute magnitudes $-22.5<r<-20$, luminosity function results from
both models using our favored parameters fall within a factor of
$\sim2$ of observed luminosity functions from SDSS.  The quenching
models also yields a qualitatively reasonable colour-magnitude
diagram when we apply a dust correction for star-forming galaxies.
Qualitatively, the red sequence grows as expected from observations,
with the most massive present-day galaxies truncating star formation
at $z>2$, and galaxies around $L^*$ continuing to evolve onto the
red sequence to $z=0$.  This leads to mean stellar ages of massive
red sequence galaxies that are roughly consistent with observations
\citep{graves07, graves09a}.
These are notable successes that show that quenching associated with
one or both of these mechanisms is on the right track towards 
understanding the evolution of massive galaxies.

Models based on quenching hot mode accretion fail to produce a red
sequence.  This is mainly because even massive galaxies obtain a
non-trivial amount of gas via cold mode accretion at the present
day~\citep{keres09_coldmode}, which is enough to keep massive
galaxies too blue.  Hence one must prevent the vast majority of
cold gas from entering into massive galaxies, be it pristine of
recycled in a wind.  Preventing hot and wind mode reaccretion
produces a red sequence, but its LF is a power law with no obvious
turnover at high masses.

Despite their broad successes, the merger and halo mass quenching
model still fail in subtle ways.  Investigating these failures
critically can help us to identify the underlying physical processes
that may be absent from our simulations.

The main failures of the merger quenching model are (a) an excess of
very bright red galaxies, (b) a shallow slope in the red sequence color-magnitude relation, and
(c) a shallow slope and a less pronounced knee in the luminosity function
(as well as the stellar mass function).

\begin{figure*}
\begin{minipage}{7in}
\centering
\includegraphics[width=5in]{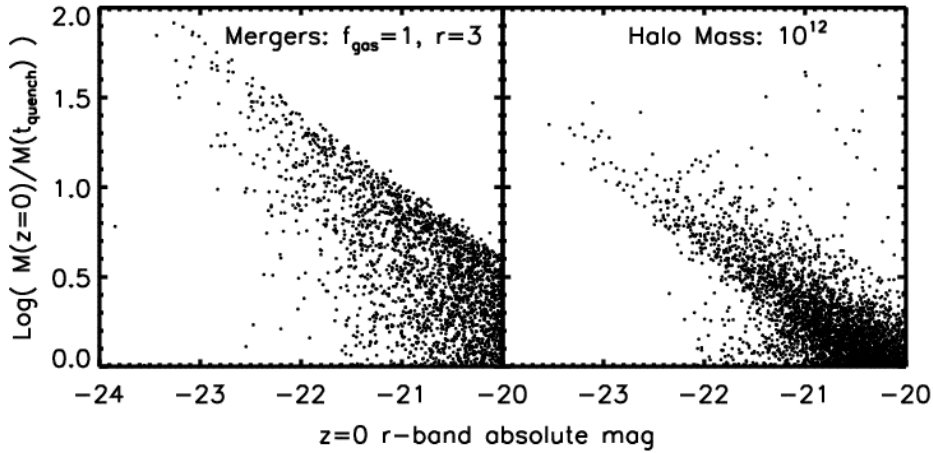}
\caption{ Mass growth of quenched galaxies via mergers.  The $y$-axis is the log of the ratio of the stellar mass of quenched galaxies at $z=0$ to the stellar mass at the time of quenching.  The upper envelope in merger quenching arises due to our resolution limit defined by a cut in stellar mass (see \S \ref{sec.merger}).   In both merger quenching and halo mass quenching, massive galaxies have obtained most of their stellar mass via mergers since they were quenched.}
\label{fig.m0_mquench}
\end{minipage}
\end{figure*}

The first problem results at least in part from post-quenching
mergers.  Quenched galaxies at the centres of massive halos build
up stellar mass by accreting satellites.  We illustrate the importance
of this stellar mass accretion in Figure \ref{fig.m0_mquench}, which
shows the $z=0$ stellar mass divided by the stellar mass at the
time of quenching as a function of $z=0$ $r$-band absolute magnitude.
Since we show only quenched galaxies, post-quenching mergers drive
all of the mass growth.  In both merger quenching and halo mass
quenching, the biggest galaxies have grown by factors $>10$ via
mergers since the time they moved onto the red sequence.  However,
the merger quenching is generally larger at a given luminosity, and
extends to factors up to $\sim 100$.

Too much mass is gained through this process, leading to excessively
bright galaxies: galaxies brighter than $r=-22.5$ ($M_* > 10^{11}
M_{\sun}$) at $z=0$ in our model have grown in mass by a median
factor of $\sim$3 since $z=1$.  This is inconsistent with observational
constraints \citep{cool08}.  Late-time mergers also explain the
presence of young stars.  Recent observations suggest no young
populations in the brightest red sequence galaxies
\citep{sanchez-blazquez09}, although up to $\sim30$\% of red sequence
galaxies do have $\sim1$--3\% of their mass in stars $<1$ Gyr old
\citep{yi05,schawinski07, kaviraj07}.  It is unclear whether the
presence and amount of young stars in our simulated quenched galaxies
is inconsistent with these data; we leave a more detailed investigation
for the future.

By building up mass with small galaxies, post-quenching mergers may
also contribute to the shallow slope of the red sequence colour-magnitude relation produced in the
merger model.  Stellar metallicities drive the slope of the red sequence,
and the mass-metallicity relations \citep{tremonti04, finlator08} tell
us that massive star-forming galaxies are more metal-rich than their less
massive counterparts.  If the mass-metallicity effect dominates over the
time evolution of metallicity at fixed mass, then a red and dead galaxy
built up through accretion of many smaller galaxies will have a lower
metallicity than that of a galaxy which grew just as massive through star
formation then moved to the red sequence.  Any accreted satellites will
have a lower metallicity, diluting the total metallicity of the galaxy.

The shape of the luminosity function is more difficult to interpret.
Its value in any luminosity bin is determined in a complicated way by
the number of galaxies moving to the red sequence from the blue cloud,
and the number of galaxies moving along the red sequence.  In principle,
decreasing the mass accreted through satellites should steepen the
red sequence slope, pushing the too-luminous galaxies to fainter $r$
magnitudes, as well as keeping their metallicities higher.

How might one lower the amount of satellite accretion?  One possibility
is that the issue is numerical.  Simulations with low resolution
tend to overmerge dynamical systems owing to stronger tidal forces
on satellites whose spatial extents are artificially enlarged by
force softening.  Another possibility is that much of the satellite
mass is actually going into diffuse halo (or intracluster) stars.
\citet{conroy07} suggests that even fairly sizeable galaxies must
be disrupted upon infall into clusters on a fairly short ($\sim
2$~Gyr) timescale.  Both trends suggest that our simulations probably
overestimate the growth of the central galaxy, at the expense of
intermediate-mass galaxies.  Unfortunately, running higher resolution
hydrodynamic simulations with sufficient volume to study very massive
galaxies is not feasible, unless one performs constrained realizations
of dense environments (in which case statistical comparisons become
more complicated).  

Turning to the halo mass quenching scenario, the main failures are
a) an excess of very bright red galaxies, b) essentially zero slope
in the red sequence, c) no knee in the luminosity function, and d)
a steep mass cut-off for blue galaxies.  The first three problems
are similar to those for merger quenching.  Very bright red galaxies
build up to an excess mass even though the satellites they accrete
have been quenched.  The problem is slightly less severe as in the
merger quenching case, likely because satellite galaxies halt their
star formation earlier.  However, this trend also causes the slightly
shallower red sequence slope as compared to merger quenching, because
the accreted satellites are less enriched.  When galaxies first
move to the red sequence, their metallicities are governed by the
mass-metallicity relation.  Since all red sequence galaxies arrive
with roughly the same mass, they all start with roughly the same
metallicity.  Subsequent mergers between pairs of red sequence
galaxies will then result in a remnant of the same metallicity.
This results in almost no trend with metallicity along the bright
red sequence ($r<-22$) in this model, as seen in
Figure~\ref{fig.red_seq_age_mass}, which is inconsistent with the
rather pronounced slope in the observed red sequence.

The last problem appears more severe and intrinsic to the halo mass
quenching scenario.  The local universe clearly contains massive,
bright blue galaxies; these are absent under this scenario.  The
difficulty is that the sharp halo mass cut produces a strong
truncation of blue galaxies above a certain stellar mass, corresponding
to the typical halo mass of truncation.  Some blue galaxies are
observed to have stellar masses well in excess of $10^{11} M_\odot$,
which under this scenario would have to live in halos of mass
$<10^{12} M_\odot$.  This is a remarkably high efficiency of star
formation from the available baryonic reservoir of such halos, well
above what is inferred from observations at any halo
mass~\citep{mcgaugh09}.  This difficulty for halo mass-based quenching
was also was found in a comparison of SAMs to SDSS data by
\citet{kimm09}.

Overall, merger quenching appears to do a reasonable job of producing
a red sequence as observed, and the discrepancies may be
traceable to numerical issues.  Of course a fully realistic model
with quenching implemented dynamically within the simulation may
alter the results, but the results here provide useful intuition
and a starting point for more sophisticated models.

How do our results compare to other galaxy evolution models?
The SAMs of \citet{croton06}, \citet{bower06}, and \citet{cattaneo06},
which use broadly similar quenching prescriptions to those we have
adopted, appear better to match the observed color-magnitude relation
of red sequence galaxies.  These models have more freedom in the
overall star formation histories of galaxies prior to quenching, since
they contain free parameters to describe the gas dynamics.  We also
note that these models tend to populate a substantial red sequence
without the additional quenchinq owing to hot haloes and/or AGN
feedback, while our models do not.  A major difference in the gas
content of haloes may result from our self-consistent treatment of
outflowing gas, which results in substantial re-accretion at late
times \citep{oppenheimer10} as compared to treatments of outflows in
these SAMs.  \citet{delucia10} point out how senstitive feedback
prescriptions are to the gas content within haloes, so this could
impact the nature and strength of quenching feedback.  We are
currently engaged in more detailed comparisons to SAMs, so we defer
further discussion to future work.

\subsection{Generic problems in massive galaxy evolution}

There are several fairly generic difficulties in understanding
quenching in the context of the observed galaxy population and its
evolution.  These are similar among all models of quenching we have
examined, and hint at an underlying failure in our understanding
or our current modeling techniques.

One issue is that there is an intrinsic tension between producing a
strong knee in the luminosity or stellar mass function, and having blue
galaxies exist in substantial numbers up to the highest stellar masses.
The former would imply a very sharp quenching of star formation at
a particular mass (stellar or halo), while the latter implies that
quenching is a much more gradual function of mass.  It is unclear how
one can resolve this tension.  Neither of our scenarios are able to do so,
with merger quenching yielding large blue galaxies, and halo mass quenching
producing a more pronounced truncation of the most massive galaxies.

A second issue is that the most massive galaxies are observed to
have grown by a surprisingly small amount over the last $\sim 10$~Gyr
\citep{cool08,banerji09}.  A consequence of this is that we overpredict
the bright end of the luminosity function in both our quenching
models; this is even the case when the satellites quench immediately
upon entering a quenched halo (in disagreement with data).  One way
to avoid this would be to have fewer or less massive satellites,
or to avoid having them fall into the central galaxy.  Perhaps more
detailed simulations will reveal that our current results are tainted
by numerical issues such as a lack of resolution, but even SAMs
have shown similar issues.  Understanding the dynamical evolution
of a collisionless stellar and dark matter halo seems to be an
unresolved problem in the context of massive galaxy evolution
since $z\sim 1$.

A final issue is one of chemical enrichment, which is perhaps more
subtle owing to current uncertainties in supernova yields and
outflows.  Nevertheless, it appears that our current simulations have
difficulty enriching massive galaxies to observed levels, and perhaps
more significantly, have trouble preferentially enriching massive red
galaxies to levels approximately twice that of $\sim L^*$ systems.
This has traditionally been a difficulty of hierarchical models that
assemble the massive end of the red sequence by dry mergers with
lower-mass systems, as this will tend to decrease the average stellar
metallicities of the most massive system.  Hence either massive
galaxies must be preferentially enriched to begin with, which is
seemingly contradictory to the ideas that they quench early on when
galaxies of a given stellar mass have lower metallicity, or else the
satellites they acquire must be quite metal-rich, which causes tension
with trying to limit the amount of late-time growth.  There is the
further problem of understanding why $\alpha$-enhancement grows with
galaxy velocity dispersion~\citep{graves07}, which again seems to
place some stringent conditions on massive galaxy
evolution~\citep{calura09, arrigoni10}.  Our current models do not
reproduce this relationship, suggesting broader difficulties with the
chemical enrichment histories of galaxies in these models.  Given the
sensitivity of the exact shape and amplitude of the red sequence to
metallicity, it is clear that a full understanding of massive galaxies
will require understanding both their star formation and chemical
enrichment histories.

\section{Summary and Conclusion} \label{sec.conclusion}

By combining cosmological SPH simulations with simple post-processing
prescriptions for the quenching of star formation, we test several
proposed quenching mechanisms associated with major mergers, halo
mass quenching, and quenching of hot and recycled wind accretion
modes.  We compared the galaxy populations resulting from these
scenarios with observations of the $z\sim0$ universe from SDSS.
With reasonable parameter choices, our merger and halo mass quenching
models vastly improve upon existing simulations by suppressing the
excessive growth of massive blue galaxies and creating a well-defined
bimodality in galaxy colors.  However, even these models fail to match
observations in detail, providing clues towards key issues in the
modeling of massive galaxy eveolution.

Our main results are:
\begin{itemize}
\item Merger quenching and halo mass quenching successfully produce
a red sequence distinct from the blue cloud, with luminosity functions
roughly consistent with those observed.
\item Quenching of hot mode accretion alone does not produce a viable
red sequence, showing that late time accretion continues via cold
mode in massive galaxies.  Further quenching wind mode accretion 
produces a red sequence, but a highly discepant luminosity function.
\item Our preferred merger quenching model rules out the re-formation of
disks after gas rich mergers, because such
re-formation leads to the build up of too many massive galaxies.
\item The halo mass quenching model is quite sensitive to the threshold
halo mass:
$M_c\approx 10^{12} M_{\sun}$ yields the best match to the red galaxy luminosity
function.
\item Both of these models yield an excess of bright red galaxies due
to mergers after the quenching process, and both yield somewhat too
few galaxies around $\sim L^*$; these problems are more severe in
the merger-based model.
\item Both models yield red sequences with too shallow slopes in the
colour-magnitude diagram, likely due to near-constant mean stellar
metallicity along the red sequence.
\item In both models, the red sequence is too blue by $\sim 0.1$ magnitudes,
likely owing to simulated galaxies being too metal-poor by $\sim \times 2$
compared to real galaxies.
\item In both models, the brightest red sequence galaxies are quenched
at redshifts $>2$, in general agreement with observed estimates.
\end{itemize}

We also identified some features that distinguish the merger quenching
model from halo mass quenching:
\begin{itemize}
\item Massive galaxies quenched via mergers include trace populations
of young stars, whereas halo mass quenched galaxies do not.  This is
true only if satellites in addition to central galaxies are quenched in the halo mass scenario, but if this
is not the case then the massive end of the LF is grossly overpopulated.
\item Merger quenching yields a population of very massive blue
galaxies, with a mass function shape similar to that of red galaxies.
Halo mass quenching creates a cutoff for blue galaxies at a
stellar mass $\sim 10^{10.5} M_{\sun}$ associated with the
$M_c=10^{12} M_{\sun}$.
\item For resolved galaxies ($M_* > 3.4 \times 10^9 M_{\sun}$), the
rate of galaxies quenched via mergers peaks at $z\approx3$, whereas
halo mass quenching occurs at an approximately constant rate.
\item The luminosity function for red sequence galaxies created via
merger quenching has a shallower (bright end) slope than that for
halo mass quenching.
\end{itemize}

Overall, the merger quenching model seems to fare somewhat better
than halo mass quenching, particularly in terms of the blue galaxy
mass function, though it still has significant discrepancies versus
data.  The generic failures in both models in terms of the
shallow red sequence slope, the lack of a pronounced knee at $L^*$,
and the excess of massive galaxies reflect that there may be generic
failures of the manner in which we quench galaxies here.  It is
possible that some of the issues are numerical, but it could be
that the quenching feedback processes actually impact subsequent
galaxy evolution in a manner not accounted for by the post-processing
technique we employ here.

These results give us general intuition about how the red sequence
is formed, as well as highlight the key issues that must be tackled
by models.  A concerted effort along both observational and theoretical
fronts will be required to fully decipher the implications of these
various trends.  In future work, we plan to implement physical
models for quenching within simulations dynamically.  By identifying
halos and merger events on the fly in our simulations, and quenching
as appropriate, we can realistically track the feedback from quenching
that influences subsequent star formation in galaxies.  Given the
evidence of the present study we favor quenching mechanisms involving
major mergers, but this is unlikely to be the entire story.  This
work is but an early step towards understanding the origin and
evolution of the fundamental bimodality in today's galaxy population.


 
 

\section*{Acknowledgments}

We would like to thank Lars Hernquist, Dusan Keres, Neal Katz, Tom Quinn,
Rodger Thompson, David Weinberg, and Ann Zabludoff for interesting and
useful discussions.

Support for this work was provided by NASA through Hubble Fellowship
grant HF-51254.01 awarded by the Space Telescope Science Institute,
which is operated by the Association of Universities for Research in
Astronomy, Inc., for NASA, under contract NAS5-26555.  Funding for
the Sloan Digital Sky Survey (SDSS) has been provided by the Alfred
P. Sloan Foundation, the Participating Institutions, the National
Aeronautics and Space Administration, the National Science Foundation,
the U.S. Department of Energy, the Japanese Monbukagakusho, and
the Max Planck Society. The SDSS Web site is http://www.sdss.org/.
The SDSS is managed by the Astrophysical Research Consortium (ARC)
for the Participating Institutions. The Participating Institutions are
The University of Chicago, Fermilab, the Institute for Advanced Study,
the Japan Participation Group, The Johns Hopkins University, Los Alamos
National Laboratory, the Max-Planck-Institute for Astronomy (MPIA), the
Max-Planck-Institute for Astrophysics (MPA), New Mexico State University,
University of Pittsburgh, Princeton University, the United States Naval
Observatory, and the University of Washington.

\bibliographystyle{mn2e} 

\bibliography{paper}


\label{lastpage}

\end{document}